\documentclass[%
reprint,
showkeys,
nofootinbib,
amsmath,amssymb,
aps,
prd,
]{revtex4-1}
\usepackage[marginal]{footmisc}
\usepackage{graphicx}
\usepackage{dcolumn}
\usepackage{bm}
\usepackage[colorlinks,allcolors=black,CJKbookmarks=True]{hyperref}
\usepackage{color}

\newcommand{\bea}{\begin{eqnarray}}
\newcommand{\ena}{\end{eqnarray}}
\newcommand{\PP}{{\cal P}_{\cal R}}
\newcommand{\PH}{{\cal P}_h}

\begin{document}
	
	\title{Enhanced curvature perturbations from spherical domain walls nucleated during inflation}
	
	\author{ Zhen-Min Zeng$^{1,2}$}
	\email{cengzhenmin@itp.ac.cn}
	
	\author{Jing Liu$^{3,4}$}
	\email{liujing@ucas.ac.cn}

	\author{Zong-Kuan Guo$^{1,2,5}$}
	\email{guozk@itp.ac.cn}

	\affiliation{$^{1}$CAS Key Laboratory of Theoretical Physics, Institute of Theoretical Physics, Chinese Academy of Sciences, Beijing 100190, China}
	\affiliation{$^{2}$School of Physical Sciences, University of Chinese Academy of Sciences, Beijing 100049, China }
	\affiliation{$^{3}$International Centre for Theoretical Physics Asia-Pacific, University of Chinese Academy of Sciences, Beijing, China }
	\affiliation{$^{4}$Taiji Laboratory for Gravitational Wave Universe (Beijing/Hangzhou), University of Chinese Academy of Sciences, Beijing, China }
	\affiliation{$^{5}$School of Fundamental Physics and Mathematical Sciences, Hangzhou Institute for Advanced Study, University of Chinese Academy of Sciences, Hangzhou 310024, China }
	
	\begin{abstract}
		We investigate spherical domain walls~(DWs) nucleated via quantum tunneling in multifield inflationary models and curvature perturbations induced by the inhomogeneous distribution of those DWs. We consider the case that the Euclidean action $S_{E}$ of DWs changes with time during inflation so that most of DWs nucleate when $S_{E}$ reaches the minimum value and the radii of DWs are almost the same.  When the Hubble horizon scale exceeds the DW radius after inflation, DWs begin to annihilate and release their energy into background radiation. Because of the random nature of the nucleation process, the statistics of DWs is of the Poisson type and the power spectrum of curvature perturbations has a characteristic slope ${\cal P}_{\cal R}(k)\propto k^{3}$. The amplitude of ${\cal P}_{\cal R}(k)$ depends on the tension and abundance of DWs at the annihilation time while the peak mode depends on the mean separation of DWs. We also numerically obtain the energy spectra of scalar-induced gravitational waves from predicted curvature perturbations which are expected to be observed in multiband gravitational-wave detectors.
		
	\end{abstract}
	\maketitle

	\section{Introduction}
	Domain walls~(DWs) are sheet-like topological defects in three spatial dimensions that can be generated in the early Universe when a discrete symmetry is spontaneously broken. A variety of new physics models predict the existance of DWs~\cite{Zeldovich:1974uw,kibble_topology_1976,PhysRevD.23.852}, such as axion models~\cite{linde_axionic_1990,PhysRevLett.48.1156,Hiramatsu:2012sc,Cicoli:2012sz,Arvanitaki:2009fg}, suppersymmetric models~\cite{Dvali:1996xe,Kovner:1997ca,takahashi_gravitational_2008,Dine:2010eb}, and the Standard model Higgs~\cite{Buttazzo:2013uya,Andreassen:2014gha,Krajewski:2016vbr}. DWs receive extensive investigation since the formation and evolution of DWs leave trace on various cosmological observations including the large-scale structure~\cite{Vilenkin:1984ib,Hill:1988vm}, the cosmic microwave background (CMB)~\cite{Zeldovich:1974uw,Takahashi:2020tqv,Gonzalez:2022mcx}, stochastic gravitational wave backgrounds~(SGWBs)~\cite{Hiramatsu:2013qaa,Saikawa:2017hiv,hiramatsu_axion_2013,Wei:2022poh,PhysRevD.104.043005,Liu:2020mru} and first-order phase transitions~\cite{Blasi:2022woz}.
	
	The formation of the DW network was regarded as a disaster in cosmology~\cite{Zeldovich:1974uw,vilenkin_cosmic_1985,saikawa_review_2017}. The curvature radius of DWs is comparable to the Hubble horizon size and the proportion of two different vacua are comparable to each other, which is well-known as the scaling behavior of DWs~\cite{1989ApJ...347..590P,hindmarsh_analytic_1996}. Numerical results confirm that the DW energy density scales as $\rho_{DW}\propto t^{-1}$ in the matter-dominated~(MD) and radiation-dominated~(RD) eras~\cite{martins_extending_2016,leite_scaling_2011,leite_accurate_2013}. Since $\rho_{DW}$ decreases much slower than the energy density of radiation and matter, DWs will finally dominate the Universe which conflicts with the present observations~\cite{planck_collaboration_planck_2020}. The temperature fluctuations of the CMB imply that DWs with tension $\sigma> \mathcal{O}(\mathrm{MeV}^{3})$ does not exist in the Universe at pressent~\cite{Zeldovich:1974uw}.
	In general, the DW problem can be avoided by introducing a bias term in the effective potential so that DWs become unstable. In this case, the DW tension and the annihilation time can be constrained by the SGWB produced from the DW network~\cite{Hiramatsu:2013qaa,Saikawa:2017hiv}, see Ref.~\cite{Jiang:2022svq} and Refs.~\cite{Bian:2022qbh,Ferreira:2022zzo} for corresponding constraints from LIGO-Virgo and pulsar timing array experiments. Ref.~\cite{Ramberg:2022irf} obtains the constraint on DWs from CMB spectral distortions.
	
	In this work, we focus on spherical DWs nucleated through quantum tunneling during inflation~\cite{basu_quantum_1991,liu_primordial_2020}. This scenario of DW formation and evolution is remarkably different from the scaling case. The radius of spherical DWs is comparable to the Hubble scale at the nucleation time. Once DWs are nucleated, they are stretched by inflation and remain stable at superhorizon scales. The Hubble horizon expands after inflation, and DWs begin to collapse when they reenter the Hubble horizon. Since the tunneling rate is exponentially suppressed by the Euclidean action of DWs, $S_{E}$, the DW problem is naturally avoided in this scenario. We consider the case that DWs have nonnegligible interaction with the matter fields so that the energy stored in DWs finally transforms into background radiation~\cite{Vachaspati:1984yi,Pujolas:2022qvs,Blasi:2022ayo}, rather than primordial black holes~(PBHs)~\cite{Tanahashi:2014sma,Garriga:2015fdk,Deng:2016vzb,Liu:2019lul,Ge:2019ihf}. According to Birkhoff's theorem, the collapse of a single spherical DW cannot produce gravitational waves (GWs). However, the inhomogeneous distribution of spherical DWs induces curvature perturbations which can serve as the source of scalar-induced GWs, providing an opportunity to verify or give constraints to this scenario. Since the nucleation of different DWs is independent of each other, the statistics of DWs obey the Poisson distribution and induce superhorizon curvature perturbations with a typical $k^{3}$ slope in the infrared power spectrum. We obtain the energy spectrum of scalar-induced GWs which is expected to be detected in multiband GW detectors. For convenience, we choose $c=8\pi G=1$ throughout this paper.
	

	
	\section{Statistical properties of DWs}
	\subsection{Nucleation of DWs via quantum tunneling}
	The nucleation of quantum topological defects during inflation is investigated in Ref.~\cite{basu_quantum_1991}, where the authors obtain the nucleation rate of spherical DWs and cosmic string loops in a de Sitter background spacetime. The Euclideanized de Sitter space is a four-sphere of radius $H^{-1}$, and DWs nucleated during inflation by quantum tunneling can be described as a three-sphere with radius $H^{-1}$. The Euclidean action is proportional to the surface area
	\begin{equation}
		S_{E}(t)=2\pi^2\sigma(t)H^{-3}(t)\,,
	\end{equation}
	where $\sigma(t)$ is the tension of DWs. The nucleation rate per unit physical volume per unit time is
	\begin{equation}
		\lambda(t)=H^{4}(t)Ae^{-S_{E}(t)}\,,
	\end{equation}
	which is obtained in the semiclassical approximation, i.e., $\sigma> H^{3}$. The nucleation rate is exponentially suppressed in the case of $\sigma\gg H^3$. However, on the contrary, the case $\sigma\ll H^3$ leads to the formation of the DW network. Thus, we mainly consider the case that $\sigma$ and $H^{3}$ are of the same order. Here $A$ is a slowly varying function of $\sigma H^{-3}$ which can be estimated as $A\sim 1~$\cite{basu_quantum_1991,garriga_nucleation_1994}. Then the number density of DWs is obtained as
	\begin{equation}
		dN=\lambda(t_{*})a^{3}(t_{*})d^3x\,dt_{*},
	\end{equation}
	where $t_{*}$ denotes the nucleation time.
	One can find that the nucleation rate of DWs is totally described by the dynamics of inflation and the evolution of the tension $\sigma(t)$. We investigate the case where the DW tension is not a constant during inflation. Since the nucleation rate is exponentially suppressed by $S_{E}$, the nucleation of DWs happens in a small period around the time when $S_{E}$ reaches the minimum so that the radii of spherical DWs are almost the same, see Ref.~\cite{liu_primordial_2020} for a specific example. In this case, the probability of nucleating a spherical DW  in a Hubble-sized region at $t_{*}$ is obtained by
	\bea
		p &=& \frac{4\pi}{3}\left(\frac{1}{H(t_{*})}\right)^3\int \frac{dN}{d^3x} \nonumber \\
          &\backsimeq & \frac{4\pi}{3} H(t_{*})e^{-S_{E}(t_{*})}\Delta t_{*}\,,
	\ena
	where $\Delta t_{*}$ is the typical time scale of the nucleation process.
	
	\subsection{Statistical distribution of DWs}
	
	The previous section indicates that spherical DWs with the comoving radius $R_{0}\sim a^{-1}(t_{*})H^{-1}(t_{*})\equiv \mathcal{H}^{-1}(t_{*})$ are randomly generated in the Universe when the Euclidean action $S_{E}$ reaches the minimum at $t_{*}$. The nucleation of DWs is irrelevant in each Hubble volume, which means that DWs satisfy the Poisson distribution. Consider a comoving volume of $(2L)^3$, where $L=n R_{0}$ and $n\gg 1$.
	To investigate the statistical properties of spherical DWs, $L$ should be larger than their comoving mean separation, $S=R_0\,p^{-1/3}$, so that plenty enough spherical DWs are contained in the volume. Let $p$ denote the probability that a spherical DW presents in a Hubble horizon and $X_i$ denote the number of spherical DWs contained in the $i$-th Hubble volume, where $i\leq n^{3}$ and $X_{i}=0$ or $1$ by definition.
	
	
	The expectation value and the varience of the random variable $X_{i}$ are respectively $E(X_i)=p$ and $D(X_i)=p(1-p)$.  Since the DW number in the volume $(2L)^{3}$ is much larger than one, according to the central limit theorem, the total DW number $X=\Sigma_{1}^{n^3}X_i$ in the comoving volume $(2L)^{3}$ is subject to Gaussian with the expectation value $E(X)=n^3 E(X_i)$ and the variance $D(X)=n^3 D(X_{i})$. We then obtain the power spectrum of curvature perturbations induced by the inhomogeneous distribution of DWs in the following.
	
	We focus on density perturbations smoothed at the scale $L$ to avoid the nonlinear effect~\cite{Bardeen:1985tr}
	\begin{equation}
		\delta(\bm{r};L)=\int\frac{d^3r^{\prime}}{(2\pi L^2)^{3/2}}\exp{\left(-\frac{|\bm{r}-\bm{r^{\prime}}|^2}{2L^2}\right)}\delta(\bm{r^{\prime}})\,,
	\end{equation}
	where $\delta(\bm{r})\equiv \delta\rho(\bm{r})/\rho$ with $\rho$ and $\delta\rho(\bm{r})$ being the spatial averaged energy density and its perturbations. Here we have chosen the Gaussian window function $\exp\left(-\frac{|\bm{r}-\bm{r^{\prime}}|^2}{2L^2}\right)$. The Fourier transformation of $\delta(\bm{r};L)$ is
	\bea
		\delta_{\bm{k}}(L) &=& \int\frac{d^3 \bm{r}}{(2\pi)^{3/2}}\delta(\bm{r};L)e^{-i\bm{k}\cdot\bm{r}} \nonumber \\
                               &=&\delta_{\bm{k}}\exp{(-k^2L^2/2)}\,,
	\ena
	where $\delta_{\bm{k}}$ is the Fourier transformation of $\delta(\bm{r})$.
	The variance of density perturbations can also be smoothed at this scale
	\bea
		\sigma^2_{\delta}(L) &=& \langle\delta(\bm{r};L)\delta(\bm{r^{\prime}};L)\rangle|_{r=r^{\prime}}=\int d\ln{k}\frac{k^3}{2\pi^2}|\delta_{k}(L)|^2 \nonumber \\
&=& \int d\ln{k}\,P_{\delta}(k)\exp{(-k^2L^2)}\,,
    \label{eq:varience}
	\ena
	where $P_{\delta}(k)\equiv \frac{k^{3}}{2\pi^{2}}|\delta_{k}|^{2}$ is the power spectrum of density perturbations.
	Assuming $P_{\delta}(k)$ has a power-law form, $P_{\delta}(k)\propto k^{n}$, then Eq.~\eqref{eq:varience} implies that the smoothed variance satisfies
	\begin{equation}
		\label{sigmap}
		\sigma^2_{\delta}(L)\propto L^{-n}.
	\end{equation}
	Total density perturbations are
	\begin{equation}
		\delta_{tot}=\frac{\delta\rho_{r}+\delta\rho_{DW}}{\rho_{r}+\rho_{DW}}\,,
	\end{equation}
	where we neglect other subdominant components in the Universe, $\delta\rho_{r}$ and $\delta\rho_{DW}$ are density perturbations of radiation and DWs, respectively. Note that $\rho_{DW}$ should be much smaller than $\rho_{r}$, otherwise DWs will dominate the Universe which conflicts with the observations. Density perturbations from radiation and DWs both contribute to total density perturbations. In general, $\delta\rho_{r}$ comes from vacuum fluctuations during inflation so that  $\delta\rho_{r}/\rho_{r}\sim 10^{-5}$, while $\delta\rho_{DW}$ comes from the random distribution of spherical DWs which could be much larger than $10^{-5}$. In the case of $\delta\rho_{DW}>\delta\rho_{r}$, curvature perturbations induced by DWs become dominated, then we have
	\bea
		\delta_{tot} &\approx& \frac{\delta\rho_{DW}}{\rho_{r}} = \frac{4\pi\sigma_{e}a^{2} R_{0}^2(X- \overline{X})}{\rho_{r}L^3 a^3} \nonumber \\
&=& \frac{\rho_{DW}}{\rho_{r}}\frac{X- \overline{X}}{ \overline{X}},
    \label{eq:deltat}
	\ena
	where $\sigma_{e}$ is the tension of DWs at the annihilation time, $\overline{X}$ is the averaged number of spherical DWs over each region of volume $(2L)^{3}$ and we have used $\rho_{DW}L^3a^3=4\pi\sigma_{e}a^{2} R_{0}^2\,\overline{X}$. Eq.~\eqref{eq:deltat} implies that $\delta_{tot}$ is also a random variable which satisfies Gaussian distribution with zero expectation value and the variance reads
	\begin{equation}
		\sigma_{\delta_{tot}}^2=\left( \frac{\rho_{DW}}{\rho_{r}}\right)^2\frac{\sigma^2_{X}}{\langle X\rangle^2}=\left( \frac{\rho_{DW}}{\rho_{r}}\right)^2\frac{(1-p)R_{0}^{3}}{p L^{3}}\,.
		\label{dwva}
	\end{equation}
	Here, we can see that $\sigma^2_{0,\delta}(L)\propto L^{-3}$, so according to the discussion in Eq.~\eqref{sigmap}, $P_{\delta}(k)$ is proportional to $k^3$. Since the length scale of induced perturbations is larger than the Hubble radius at the annihilation time, we can safely use the superhorizon relation
	\begin{equation}
		P_{\delta}(k)=\frac{16}{81} \PP(k)\,,
		\label{dcps}
	\end{equation}
	where $\PP(k)$ is the power spectrum of curvature perturbations. Eq.~\eqref{dcps} allows us to parameterize $\PP(k)$ in the form $\PP(k)=A_{d}(k/k_{cut})^{3}$ where  $k_{cut}$ is a cutoff scale arising from  the requirement of central limit theorem $L>S$. Since in smaller scale, the distribution of DWs become nongaussian and $\PP(k)$ decrease rapidly, we simply apply the approximation $k_{cut}=S^{-1}$ and $\PP(k)=0$ for $k>k_{cut}$. Then, Eq.~\eqref{eq:varience} could be rewritten in the form
	\bea
		\sigma_{\delta_{tot}}^2(L) &=& \frac{16 A_{d}}{81 (k_{cut}L)^{3}}\int_{0}^{k_{cut}}d\left(kL\right)\exp{\left(-k^2 L^2 \right)}\left(kL\right)^{2} \nonumber \\
&=& \frac{4\sqrt{\pi} A_{d}}{81 (k_{cut}L)^{3}}\,,
	\ena
	which helps to determine the coefficient
	\begin{equation}
		A_{d}=\frac{9}{4\sqrt{\pi}}\left( \frac{\rho_{DW}}{\rho_{r}}\right)^2\frac{1-p}{p}\,.
	\end{equation}
	The final result of the power spectrum of induced curvature perturbations is
	\begin{equation}\label{eq:finalpr}
		\PP=\left\{
		\begin{aligned}
			&\frac{9}{4\sqrt{\pi}}\left( \frac{\rho_{DW}}{\rho_{r}}\right)^2\frac{1-p}{p}\left(\frac{k}{k_{cut}}\right)^3\quad &\mathrm{for}\; k\leq k_{cut}\,,\\
			&0\quad &\mathrm{for}\; k>k_{cut}\,.
		\end{aligned}
		\right.
	\end{equation}

	\subsection{Evolution of the DW energy density}
	
	At the time $t_{*}$ when DWs are nucleated, the energy density of DWs is
	\bea
		\rho_{DW}(t_{*}) &=& 4\pi\left(\frac{1}{H(t_{*})}\right)^2\sigma(t_{*})\frac{dN}{a^3(t_{*})d^3x} \nonumber \\
&=& 3H(t_{*})\sigma(t_{*})p \,.
	\ena
	Afterward, DWs are stretched by inflation and their tension evolves with time. At the end of inflation $t_{e}$
	\bea \rho_{DW} &=& 4\pi\left(\frac{a(t_{e})}{H(t_{*})a(t_{*})}\right)^2\sigma(t_{e})\frac{dN}{a^3(t_{e})d^3x} \nonumber \\
&=& 3H(t_{*})\frac{a(t_{*})}{a(t_{e})}\sigma(t_{e})p,
	\ena
	\begin{equation}
		\rho_{tot}=3 H^2(t_{e})\,.
	\end{equation}
	Here, we assume a short reheating process and the Universe quickly enters the RD era after inflation. If the tension of DWs remains constant after inflation, the energy density of spherical DWs scales as $\rho_{DW}\propto a^{-1}$(the area of a single spherical DW scales as $a^{2}$ and the number density of spherical DWs scales as $a^{-3}$) at superhorizon scales, while the total energy density scale as $\rho_{tot}\approx\rho_{r}\propto a^{-4}$ in the RD era, then we have
	\begin{equation}
		\frac{\rho_{DW}}{\rho_{r}}\bigg|_{t_{r}}=\frac{H(t_{*})}{H^2(t_{e})}\frac{a(t_{*})}{a(t_{e})}\sigma(t_{e})p\left(\frac{a(t_{r})}{a(t_{e})}\right )^3,
	\end{equation}
	 where $t_{r}$ corresponding to the time that DWs reenter the horizon(annihilation time), which is long before reenter time of $k_{cut}$. Thus, the other undetermined term in Eq.~\eqref{eq:finalpr}, $\rho_{DW}/\rho_{r}$, can be obtained from physical parameters $\sigma(t_{e})$ and $S_{E}$ during inflation. Note that $\rho_{DW}$ cannot exceed $\rho_{r}$ even inside the Hubble horizons containing a spherical DW, otherwise, the Hubble horizon collapses into a PBH before $t_{r}$, which is investigated as the ``supercritical" case in~\cite{deng_primordial_2017}. This condition requires $\rho_{DW}/\rho_{r}<p$.
	If the interaction between DWs and matter fields is nonnegligible, spherical DWs dissipate their energy into background radiation at the annihilation time. Thus, the random distribution of DWs finally leads to density perturbations in the background radiation.
	\section{Scalar-induced GWs}
	
	
	Induced curvature perturbations reenter the Hubble horizon and begin to evolve soon after the annihilation of DWs. Since the collapse of a single spherical DW cannot produce GWs, the unique SGWB in this scenario is induced by curvature perturbations predicted in the last section.
	In this section, we introduce the formula to calculate GWs induced by scalar perturbations at the second order~\cite{Espinosa:2018eve,Kohri:2018awv}. The perturbed  metric of a Friedmann-Robertson-Walker Universe in the Newtonian gauge reads
	\bea
		ds^2 &=& a^2(\tau)\Big\{-(1+2\Phi)d\tau^2 \nonumber \\
&& +[(1-2\Psi)\delta_{ij} + \frac{1}{2} h_{ij}]dx^idx^j\Big\},
	\ena
	where $\tau$ is conformal time, $\Phi$ and $\Psi$ represent scalar perturbations and $h_{ij}$ denotes tensor perturbations of the second order. Here, we neglect vector perturbations and first-order tensor perturbations. We also neglect the anisotropic pressure so that we take $\Phi=\Psi$ in the following. The equation of motion~(EoM) of the tensor modes sourced by curvature perturbations reads
	\begin{equation}
		h_{ij}^{\prime\prime}+2\mathcal{H}h_{ij}^{\prime}-\nabla^2h_{ij}=-4\Pi_{ij}^{lm}\mathcal{S}_{lm},
	\end{equation}
	$\Pi_{ij}^{lm}$ is the transverse-traceless projection operator and $S_{lm}$ is the scalar-induced source term. Tensor perturbations can be expanded into the Fourier modes as
	\begin{equation}
		h_{ij}(\tau,\bm{x})=\int\frac{d^3k}{(2\pi)^{3/2}}\left[e_{ij}^{+}(\bm{k})h_{\bm{k}}^{+}+e_{ij}^{\times}(\bm{k})h_{\bm{k}}^{\times}\right]e^{i\bm{k}\cdot\bm{x}},
	\end{equation}
	where $e_{ij}^{\lambda}(\bm{k})(\lambda=+,\times)$ are the polarization tensors.  Similarly, the Fourier modes of the source term are
	\begin{equation}
		\Pi_{ij}^{lm}\mathcal{S}_{lm}(\tau,\bm{x})=\sum_{\lambda=+,\times}\int\frac{d^3 k}{(2\pi)^{3/2}}e_{ij}^{\lambda}(\bm{k})e^{\lambda,lm}(\bm{k})S_{lm}(\tau,k)\,.
	\end{equation}
	Then, the EoM of the tensor modes $h_{\bm{k}}(\tau)$ can be written in the form
	\begin{equation}\label{eq:hkSk}
		h_{\bm{k}}^{\prime\prime}(\tau)+2\mathcal{H}h_{\bm{k}}^{\prime}(\tau)+k^2h_{\bm{k}}(\tau)=4S_{\bm{k}}(\tau)\,.
	\end{equation}
	Here, we ignore the upper index of two different polarization modes since they satisfy the same equation. The source term $S_{\bm{k}}$ reads
	\bea
		S_{\bm{k}}(\tau) &=& \int\frac{d^3 q}{(2\pi)^{3/2}}e_{ij}(\bm{k})q_{i}q_{j}\Bigg[ 2\Phi_{\bm{q}}\Phi_{\bm{k-q}}+\frac{4}{3(1+3\omega)} \nonumber \\
&& \times (\mathcal{H}^{-1}\Phi^{\prime}_{\bm{q}}+\Phi_{\bm{q}})(\mathcal{H}^{-1}\Phi^{\prime}_{\bm{k-q}}+\Phi_{\bm{k-q}})\Bigg]\,,
	\ena
	where $\omega$ is the equation of state parameter of the Universe and $\omega=1/3$ in the RD era. The Newtonian potential $\Phi$ obeys the following equation
	\begin{equation}
		\Phi_{k}^{\prime\prime}+\frac{6(1+\omega)}{1+3\omega}\frac{1}{\tau}\Phi_{k}^{\prime}+\omega^2k^2\Phi_{k}=0\,,
	\end{equation}
	where we ignore entropy perturbations. The initial value of the Newtonian potential, $\Phi_{k,0}$, is related to the power spectrum of curvature perturbations as
	\begin{equation}
		\left\langle\Phi_{k,0}\Phi_{k^{\prime},0}\right\rangle=\delta^{(3)}(\bm{k}-\bm{k^{\prime}})\frac{2\pi^2}{k^3}\left(\frac{3+3\omega}{5+3\omega}\right)^2\PP(k)\,.
	\end{equation}
	We can use the Green's function method to solve Eq.~\eqref{eq:hkSk}
	\begin{equation}
		a(\tau) h_{\bm{k}}(\tau)=4\int^{\tau}d\tau_{1}\mathcal{G}(\tau,\tau_{1})a(\tau_{1})S_{\bm{k}}(\tau_{1})\,,
	\end{equation}
	where the Green function $\mathcal{G}(\tau,\tau^{\prime})$ is the solution of
	\begin{equation}
		\mathcal{G}^{\prime\prime}_{\bm{k}}(\tau,\tau_{1})+\left(k^2-\frac{a^{\prime\prime}(\tau)}{a(\tau)}\right)\mathcal{G}_{\bm{k}}(\tau,\tau_{1})=\delta(\tau-\tau_{1})\,.
	\end{equation}
	The power spectrum of tensor perturbations is defined by
	\begin{equation}
		\langle h_{\bm{k}}^{\lambda}(\tau)h_{\bm{k}^{\prime}}^{\lambda^{\prime}}(\tau)\rangle=\delta_{\lambda\lambda^{\prime}}\delta^{(3)}(\bm{k}-\bm{k}^{\prime})\frac{2\pi^2}{k^3}\PH(\tau,k)\,.
	\end{equation}
	The energy spectrum of GWs is defined as
	\begin{equation}
		\Omega_{\mathrm{GW}}(\tau,k)\equiv\frac{1}{\rho_{tot}}\frac{d\rho_{\mathrm{GW}}}{d\ln k}=\frac{1}{24}\left(\frac{k}{\mathcal{H}(\tau)}\right)^{2}\overline{\PH(\tau,k)}\,,
	\end{equation}
	where the overline represents the oscillation average and the two polarization modes have been added up.
	Then, in the RD era, $\Omega_{\mathrm{GW}}$ of scalar-induced GWs can be evaluated by the following integral
		\bea
\Omega_{\mathrm{GW}}(\tau,k)&=&\frac{1}{12}\int_{0}^{\infty}dv\int_{|1-v|}^{|1+v|}du\left(\frac{4v^2-(1+v^2-u^2)^2}{4uv}\right)^2 \nonumber \\
  && \times \PP(ku)\PP(kv) \left(\frac{3}{4u^3v^3}\right)^2(u^2+v^2-3)^2 \nonumber \\
  && \times \Bigg\{\Bigg[-4uv+(u^2+v^2-3)\ln{\left|\frac{3-(u+v)^2}{3-(u-v)^2}\right|}\Bigg]^2 \nonumber \\
  && +\pi^2(u^2+v^2-3)^2\Theta(u+v-\sqrt{3})\Bigg\}\,.
		\ena
	In order to obtain the GW energy spectrum at present, we need to take the thermal history into consideration
	\begin{equation}
		\Omega_{\mathrm{GW}}(\tau_0,k)=\Omega_{\gamma,0}\left(\frac{g_{*,0}}{g_{*,eq}}\right)^{\frac{1}{3}}\Omega_{\mathrm{GW}}(\tau_{eq},k)\,,
	\end{equation}
	where $\Omega_{\gamma,0}$ is the density parameter of radiation at present, $g_{*,0}$ and $g_{*,eq}$ are the effect numbers of relativistic degrees of freedom at present and the radiation-matter equality, $\tau_{eq}$.
	
\begin{figure}[ht]
		\centering
		\includegraphics[width=0.45\textwidth]{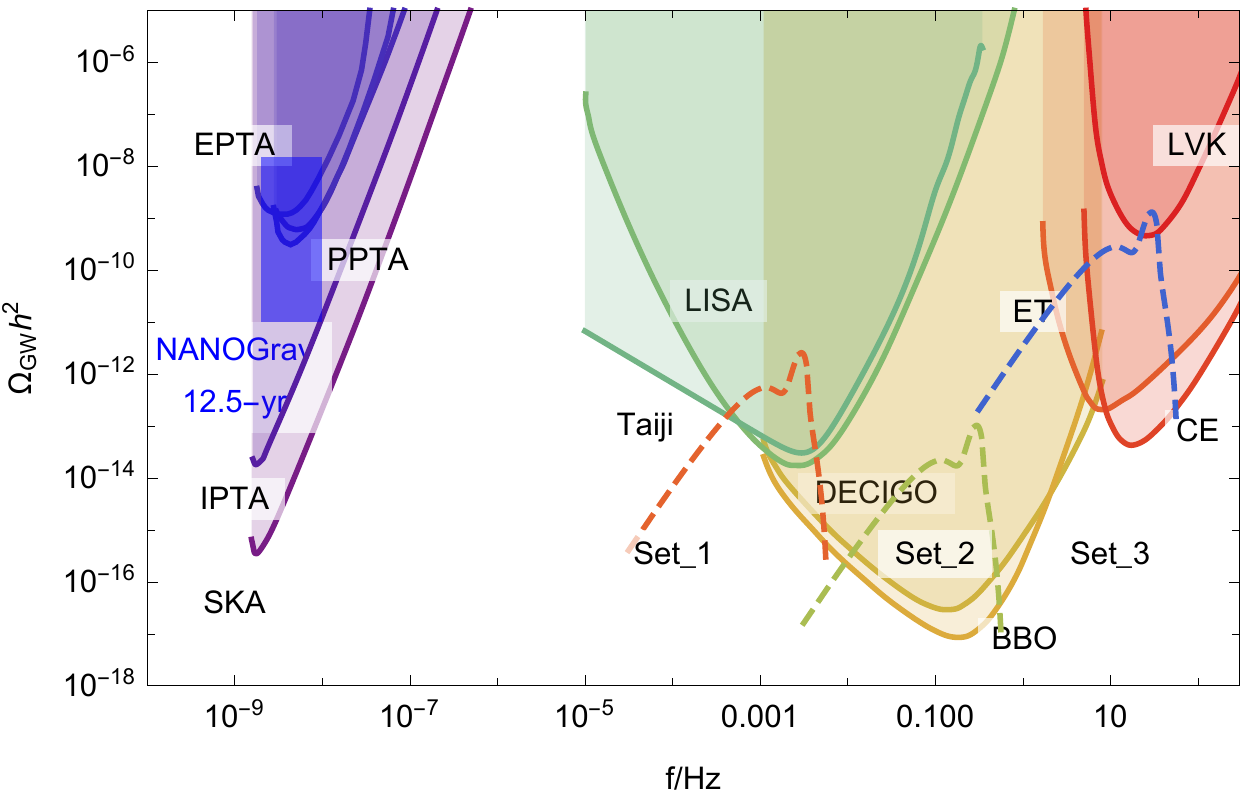}
		\caption{Predicted energy spectra of scalar-induced GWs with the parameter set 1 (red), set 2 (yellow) and set 3 (blue) in Table.~\ref{tab:my_label}, respectively, which are expected to be observed by LISA/Taiji, DECIGO/BBO and LVK/ET/CE respectively.
We show the sensitivity curves of the GW detectors, including EPTA~\cite{Lentati:2015qwp}, PPTA~\cite{Shannon:2015ect}, NANOGrav~\cite{Arzoumanian:2018saf,NANOGrav:2020bcs}, IPTA~\cite{Hobbs:2009yy}, SKA~\cite{Carilli:2004nx}, LISA~\cite{Audley:2017drz} Taiji~\cite{Guo:2018npi}, DECIGO~\cite{Kawamura:2011zz}, BBO~\cite{phinney2004big}, LIGO, Virgo and KAGRA~( LVK)~\cite{TheLIGOScientific:2014jea,Somiya:2011np}, CE~\cite{Reitze:2019iox}, ET~\cite{Punturo:2010zz}, which are summarized in Ref.~\cite{Schmitz:2020syl}.}
    \label{fig:GW}
	\end{figure}

	We choose three sets of parameters in Table.~\ref{tab:my_label} to show the predictions of $\Omega_{\mathrm{GW}}$. The probability $p$, the abundance of DWs $\rho_{DW}/\rho_{r}$ and the cutoff scale $k_{cut}$ are determined by the DW tension $\sigma$ and the evolution of $S_{E}$ during inflation, and thus can also be treated as free parameters with the only constraint $\rho_{DW}/\rho_{r}<p$ to avoid the formation of PBHs.
	For the three parameter sets, the predicted $\Omega_{\mathrm{GW}}$ peak at $0.001$Hz, $0.1$Hz and $10$Hz, respectively, which are expected to be detected by multiband GW detectors, including LISA/Taiji~(set 1), DECIGO/BBO~(set 2), CE/ET and LIGO-Virgo-KAGRA collaboration~(set 3), as shown in Fig.~\ref{fig:GW}.
	\begin{table}[ht]
		\centering
		\begin{tabular}{|c|c|c|c|}
			\hline
			set & $\rho_{DW}/\rho_{tot}$&$k_{cut}/\mathrm{Mpc}^{-1}$&$p$\\
			\hline
			1 & $2\times10^{-3}$&$2\times10^{10}$&$4\times10^{-3}$\\
			2 & $10^{-3}$&$2\times10^{12}$&$5\times10^{-3}$\\
			3 & $5\times10^{-2}$&$2\times10^{14}$&$10^{-1}$\\
			\hline
		\end{tabular}
		\caption{Parameter sets we choose in this paper.}
		\label{tab:my_label}
	\end{table}
	
	The observation of the CMB temperature anisotropies give strict constraints on primordial curvature perturbations $\PP(k)\approx 2\times10^{-9}$ for $10^{-3}\,\mathrm{Mpc}^{-1}\lesssim k\lesssim1\,\mathrm{Mpc}^{-1}$. At smaller scales, the observations of CMB spectral distortions, big-bang nucleosynthesis and ultracompact minihaloes also give constraints on $\PP(k)$ at the scales of $1\,\mathrm{Mpc}^{-1}\lesssim k\lesssim 10^{8}\,\mathrm{Mpc}^{-1}$~\cite{Emami:2017fiy,Gow:2020bzo}.
	In the three parameter sets of Table.~\ref{tab:my_label}, the results of $\PP$ are orders of magnitude smaller than $10^{-10}$ at the scale $k\sim 3\times 10^{7}\,\mathrm{Mpc}^{-1}$ so that we safely avoid the constraints on $\PP(k)$ from the observations of the CMB spectrum distortion and the ultracompact minihalo abundance. However, because of the limit on $\PP(k)$, GWs are constrained to be $\Omega_{\mathrm{GW}}\lesssim 10^{-17}$ which is too weak to be observed in the nanohertz band by SKA.
	
	\section{Realistic example}
	We show the results of $\PP(k)$ and $\Omega_{\mathrm{GW}}$ of scalar-induced GWs in a realistic two-field inflationary model where $S_{E}$ changes with time during inflation.
	The action reads
	\begin{equation}
		S=\int d^4x\sqrt{-g}\left[-\frac{R}{2}+
		\frac{1}{2}\partial^{\mu}\phi\partial_{\mu}\phi+ \frac{1}{2}\partial^{\mu}\chi\partial_{\mu}\chi+V(\phi,\chi)\right],
	\end{equation}
	with the potential $V(\phi,\chi)$
	\begin{equation}\label{eq:pot}
		V(\phi,\chi)=\frac{\lambda_{\chi}}{4}\left[\chi^2-\alpha^2(\phi-\phi_c)^2-m^2\right]^2+f(\phi)\,,
	\end{equation}
	which provides two degenerate vacua in the $\chi$ direction. Since the dynamics of $\phi$ is unaffected by $\chi$ during inflation, the term $f(\phi)$ alone is responsible for the inflationary dynamics and generating primordial perturbations~\cite{Planck:2018vyg}.
	
\begin{figure}[ht]
		\centering
		\includegraphics[width=0.45\textwidth]{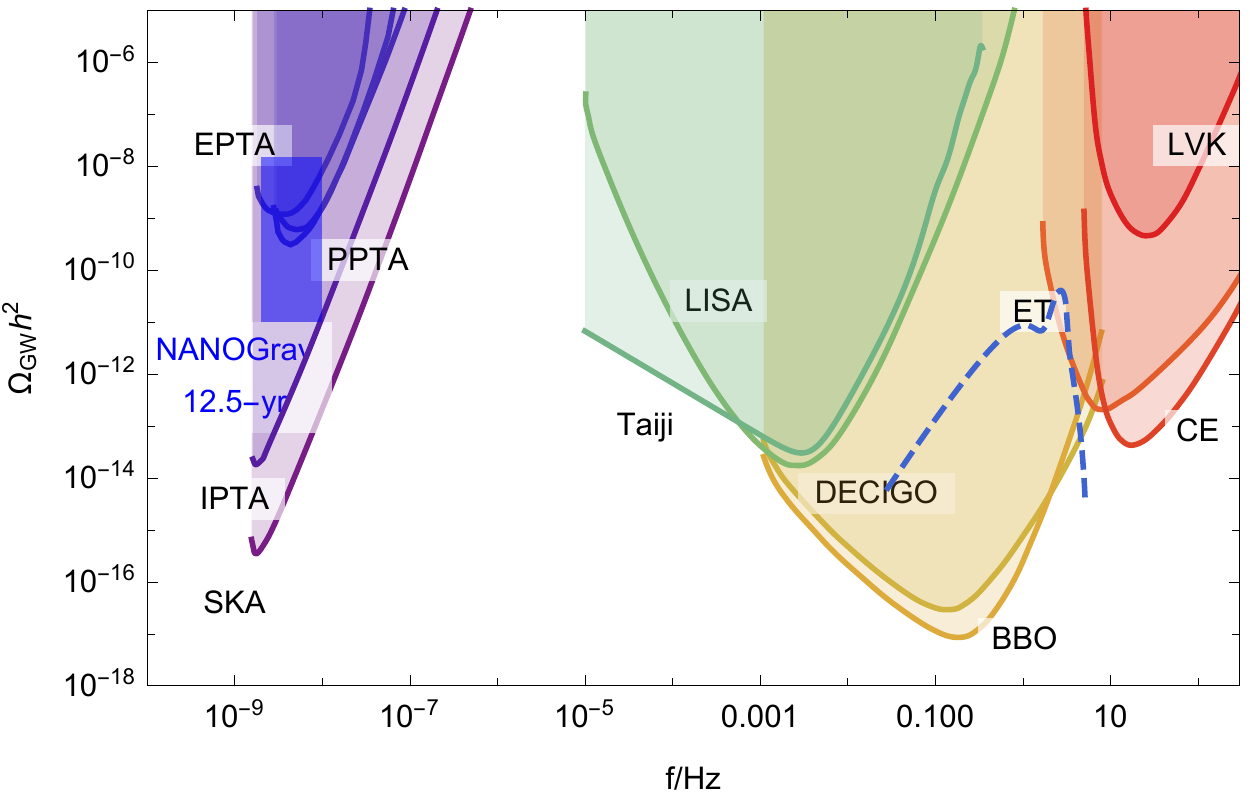}
		\caption{Predicted energy spectrum (dashed blue) of scalar-induced GWs as a specific realization of our mechanism.}
		\label{fig:s4}
	\end{figure}

	The Friedmann equation and the EoMs of $\phi$ and $\chi$ are
	\bea
			&& H^2=\frac{1}{3}\left[\frac{1}{2}\dot{\phi}^2+\frac{1}{2}\dot{\chi}^2+V(\phi,\chi)\right]\,,\nonumber \\
			&& \ddot{\phi}+3H\dot{\phi}+\frac{\partial V}{\partial \phi}=0\,, \nonumber\\
			&& \ddot{\chi}+3H\dot{\chi}+\frac{\partial V}{\partial \chi}=0\,.
	\ena
	We consider Starobinsky inflation with the potential
	\begin{equation}
		f(\phi)=\Lambda_0\left(1-e^{-\sqrt{\frac{2}{3}}\phi}\right)^2\,.
	\end{equation}
	The tension of DWs is a function of $\phi(t)$
	\begin{equation}
		\sigma_{\chi}(t)=\frac{4}{4}\sqrt{\frac{\lambda_{\chi}}{2}}\left[\alpha^2(\phi^2-\phi_{c}^2)^2+m^2\right]^{\frac{3}{2}}.
	\end{equation}
	At the time $\phi(t)=\phi_{c}$, the Euclidean action  $S_{E}=2\pi^2\sigma_{\chi}(t)H^{-3}(t)$ reaches minimum and most of DWs nucleate. We choose a specific parameter set to show the result of the energy spectrum of induced GWs in Fig.~\ref{fig:s4}, where $\lambda_{\chi}=0.3,\alpha=1\times 10^{-5},\phi_{c}=3.9,m=5\times10^{-6}$.
	The initial value of the scalar fields are set to be $\phi_{i}=5.1$and $\chi_{i}=0.0008$ so that the predicted $e$-folds is $N=50$. $\Omega_{\mathrm{GW}}$ peaks at about $1$Hz with the peak value $\sim 10^{-10}$, which is expected to be observed by DECIGO and BBO.
	
	\section{Conclusion and disscussion}
	
	
	We have investigated spherical DWs nucleated during inflation via quantum tunneling and found their random distribution induces curvature perturbations at the length scales larger than the mean separation of spherical DWs. The statistics of DWs turn out to be the Poisson type and the power spectrum of induced curvature perturbations is proportional to $k^{3}$. We numerically calculate the energy spectrum of scalar-induced GWs in terms of $\PP(k)$ which can be detected by multiband GW detectors. Since the collapse of spherical DWs does not directly produce GWs, our work provides a practical method to detect or constrain the case that the energy of spherical DWs decays into radiation.
	
	This result is also applicable to false vacuum bubbles nucleated during inflation, proposed in Refs.~\cite{Ashoorioon:2015hya,Ashoorioon:2020hln,Deng:2017uwc,Deng:2020mds}, where the vacua of the effective potential are nondegenerate. Induced curvature perturbations from Poisson distribution have been also discussed in other physical processes in the early Universe such as PBH formation~\cite{Papanikolaou:2020qtd,Papanikolaou:2021uhe,Bhaumik:2022pil,Bhaumik:2022zdd,Domenech:2020ssp} and first-order phase transitions~\cite{Liu:2022lvz}. These processes directly produce another SGWB from the transverse-traceless part of the energy-momentum tensor, which could be distinguished from the random distribution of spherical DWs or false vacuum bubbles.
	
	\section{Acknowledgement}
	This work is supported in part by the National Key Research and Development Program of China Grant No.2020YFC2201501,
in part by the National Natural Science Foundation of China Grants No. 12105060, No. 12147103, No. 12075297 and No. 12235019,
and in part by the Fundamental Research Funds for the Central Universities.

	\bibliographystyle{apsrev4-2}
	\bibliography{dwinducebib}

\begin{thebibliography}{81}%
\makeatletter
\providecommand \@ifxundefined [1]{%
 \@ifx{#1\undefined}
}%
\providecommand \@ifnum [1]{%
 \ifnum #1\expandafter \@firstoftwo
 \else \expandafter \@secondoftwo
 \fi
}%
\providecommand \@ifx [1]{%
 \ifx #1\expandafter \@firstoftwo
 \else \expandafter \@secondoftwo
 \fi
}%
\providecommand \natexlab [1]{#1}%
\providecommand \enquote  [1]{``#1''}%
\providecommand \bibnamefont  [1]{#1}%
\providecommand \bibfnamefont [1]{#1}%
\providecommand \citenamefont [1]{#1}%
\providecommand \href@noop [0]{\@secondoftwo}%
\providecommand \href [0]{\begingroup \@sanitize@url \@href}%
\providecommand \@href[1]{\@@startlink{#1}\@@href}%
\providecommand \@@href[1]{\endgroup#1\@@endlink}%
\providecommand \@sanitize@url [0]{\catcode `\\12\catcode `\$12\catcode
  `\&12\catcode `\#12\catcode `\^12\catcode `\_12\catcode `\%12\relax}%
\providecommand \@@startlink[1]{}%
\providecommand \@@endlink[0]{}%
\providecommand \url  [0]{\begingroup\@sanitize@url \@url }%
\providecommand \@url [1]{\endgroup\@href {#1}{\urlprefix }}%
\providecommand \urlprefix  [0]{URL }%
\providecommand \Eprint [0]{\href }%
\providecommand \doibase [0]{https://doi.org/}%
\providecommand \selectlanguage [0]{\@gobble}%
\providecommand \bibinfo  [0]{\@secondoftwo}%
\providecommand \bibfield  [0]{\@secondoftwo}%
\providecommand \translation [1]{[#1]}%
\providecommand \BibitemOpen [0]{}%
\providecommand \bibitemStop [0]{}%
\providecommand \bibitemNoStop [0]{.\EOS\space}%
\providecommand \EOS [0]{\spacefactor3000\relax}%
\providecommand \BibitemShut  [1]{\csname bibitem#1\endcsname}%
\let\auto@bib@innerbib\@empty
\bibitem [{\citenamefont {Zeldovich}\ \emph {et~al.}(1974)\citenamefont
  {Zeldovich}, \citenamefont {Kobzarev},\ and\ \citenamefont
  {Okun}}]{Zeldovich:1974uw}%
  \BibitemOpen
  \bibfield  {author} {\bibinfo {author} {\bibfnamefont {Y.~B.}\ \bibnamefont
  {Zeldovich}}, \bibinfo {author} {\bibfnamefont {I.~Y.}\ \bibnamefont
  {Kobzarev}},\ and\ \bibinfo {author} {\bibfnamefont {L.~B.}\ \bibnamefont
  {Okun}},\ }\href@noop {} {\bibfield  {journal} {\bibinfo  {journal} {Zh.
  Eksp. Teor. Fiz.}\ }\textbf {\bibinfo {volume} {67}},\ \bibinfo {pages} {3}
  (\bibinfo {year} {1974})}\BibitemShut {NoStop}%
\bibitem [{\citenamefont {Kibble}(1976)}]{kibble_topology_1976}%
  \BibitemOpen
  \bibfield  {author} {\bibinfo {author} {\bibfnamefont {T.~W.~B.}\
  \bibnamefont {Kibble}},\ }\href {https://doi.org/10.1088/0305-4470/9/8/029}
  {\bibfield  {journal} {\bibinfo  {journal} {Journal of Physics A:
  Mathematical and General}\ }\textbf {\bibinfo {volume} {9}},\ \bibinfo
  {pages} {1387} (\bibinfo {year} {1976})}\BibitemShut {NoStop}%
\bibitem [{\citenamefont {Vilenkin}(1981)}]{PhysRevD.23.852}%
  \BibitemOpen
  \bibfield  {author} {\bibinfo {author} {\bibfnamefont {A.}~\bibnamefont
  {Vilenkin}},\ }\href {https://doi.org/10.1103/PhysRevD.23.852} {\bibfield
  {journal} {\bibinfo  {journal} {Phys. Rev. D}\ }\textbf {\bibinfo {volume}
  {23}},\ \bibinfo {pages} {852} (\bibinfo {year} {1981})}\BibitemShut
  {NoStop}%
\bibitem [{\citenamefont {Linde}\ and\ \citenamefont
  {Lyth}(1990)}]{linde_axionic_1990}%
  \BibitemOpen
  \bibfield  {author} {\bibinfo {author} {\bibfnamefont {A.~D.}\ \bibnamefont
  {Linde}}\ and\ \bibinfo {author} {\bibfnamefont {D.~H.}\ \bibnamefont
  {Lyth}},\ }\href {https://doi.org/10.1016/0370-2693(90)90613-B} {\bibfield
  {journal} {\bibinfo  {journal} {Physics Letters B}\ }\textbf {\bibinfo
  {volume} {246}},\ \bibinfo {pages} {353} (\bibinfo {year}
  {1990})}\BibitemShut {NoStop}%
\bibitem [{\citenamefont {Sikivie}(1982)}]{PhysRevLett.48.1156}%
  \BibitemOpen
  \bibfield  {author} {\bibinfo {author} {\bibfnamefont {P.}~\bibnamefont
  {Sikivie}},\ }\href {https://doi.org/10.1103/PhysRevLett.48.1156} {\bibfield
  {journal} {\bibinfo  {journal} {Phys. Rev. Lett.}\ }\textbf {\bibinfo
  {volume} {48}},\ \bibinfo {pages} {1156} (\bibinfo {year}
  {1982})}\BibitemShut {NoStop}%
\bibitem [{\citenamefont {Hiramatsu}\ \emph
  {et~al.}(2013{\natexlab{a}})\citenamefont {Hiramatsu}, \citenamefont
  {Kawasaki}, \citenamefont {Saikawa},\ and\ \citenamefont
  {Sekiguchi}}]{Hiramatsu:2012sc}%
  \BibitemOpen
  \bibfield  {author} {\bibinfo {author} {\bibfnamefont {T.}~\bibnamefont
  {Hiramatsu}}, \bibinfo {author} {\bibfnamefont {M.}~\bibnamefont {Kawasaki}},
  \bibinfo {author} {\bibfnamefont {K.}~\bibnamefont {Saikawa}},\ and\ \bibinfo
  {author} {\bibfnamefont {T.}~\bibnamefont {Sekiguchi}},\ }\href
  {https://doi.org/10.1088/1475-7516/2013/01/001} {\bibfield  {journal}
  {\bibinfo  {journal} {JCAP}\ }\textbf {\bibinfo {volume} {01}},\ \bibinfo
  {pages} {001}},\ \Eprint {https://arxiv.org/abs/1207.3166} {arXiv:1207.3166
  [hep-ph]} \BibitemShut {NoStop}%
\bibitem [{\citenamefont {Cicoli}\ \emph {et~al.}(2012)\citenamefont {Cicoli},
  \citenamefont {Goodsell},\ and\ \citenamefont {Ringwald}}]{Cicoli:2012sz}%
  \BibitemOpen
  \bibfield  {author} {\bibinfo {author} {\bibfnamefont {M.}~\bibnamefont
  {Cicoli}}, \bibinfo {author} {\bibfnamefont {M.}~\bibnamefont {Goodsell}},\
  and\ \bibinfo {author} {\bibfnamefont {A.}~\bibnamefont {Ringwald}},\ }\href
  {https://doi.org/10.1007/JHEP10(2012)146} {\bibfield  {journal} {\bibinfo
  {journal} {JHEP}\ }\textbf {\bibinfo {volume} {10}},\ \bibinfo {pages}
  {146}},\ \Eprint {https://arxiv.org/abs/1206.0819} {arXiv:1206.0819 [hep-th]}
  \BibitemShut {NoStop}%
\bibitem [{\citenamefont {Arvanitaki}\ \emph {et~al.}(2010)\citenamefont
  {Arvanitaki}, \citenamefont {Dimopoulos}, \citenamefont {Dubovsky},
  \citenamefont {Kaloper},\ and\ \citenamefont
  {March-Russell}}]{Arvanitaki:2009fg}%
  \BibitemOpen
  \bibfield  {author} {\bibinfo {author} {\bibfnamefont {A.}~\bibnamefont
  {Arvanitaki}}, \bibinfo {author} {\bibfnamefont {S.}~\bibnamefont
  {Dimopoulos}}, \bibinfo {author} {\bibfnamefont {S.}~\bibnamefont
  {Dubovsky}}, \bibinfo {author} {\bibfnamefont {N.}~\bibnamefont {Kaloper}},\
  and\ \bibinfo {author} {\bibfnamefont {J.}~\bibnamefont {March-Russell}},\
  }\href {https://doi.org/10.1103/PhysRevD.81.123530} {\bibfield  {journal}
  {\bibinfo  {journal} {Phys. Rev. D}\ }\textbf {\bibinfo {volume} {81}},\
  \bibinfo {pages} {123530} (\bibinfo {year} {2010})},\ \Eprint
  {https://arxiv.org/abs/0905.4720} {arXiv:0905.4720 [hep-th]} \BibitemShut
  {NoStop}%
\bibitem [{\citenamefont {Dvali}\ and\ \citenamefont
  {Shifman}(1997)}]{Dvali:1996xe}%
  \BibitemOpen
  \bibfield  {author} {\bibinfo {author} {\bibfnamefont {G.~R.}\ \bibnamefont
  {Dvali}}\ and\ \bibinfo {author} {\bibfnamefont {M.~A.}\ \bibnamefont
  {Shifman}},\ }\href {https://doi.org/10.1016/S0370-2693(97)00131-7}
  {\bibfield  {journal} {\bibinfo  {journal} {Phys. Lett. B}\ }\textbf
  {\bibinfo {volume} {396}},\ \bibinfo {pages} {64} (\bibinfo {year} {1997})},\
  \bibinfo {note} {[Erratum: Phys.Lett.B 407, 452 (1997)]},\ \Eprint
  {https://arxiv.org/abs/hep-th/9612128} {arXiv:hep-th/9612128} \BibitemShut
  {NoStop}%
\bibitem [{\citenamefont {Kovner}\ \emph {et~al.}(1997)\citenamefont {Kovner},
  \citenamefont {Shifman},\ and\ \citenamefont {Smilga}}]{Kovner:1997ca}%
  \BibitemOpen
  \bibfield  {author} {\bibinfo {author} {\bibfnamefont {A.}~\bibnamefont
  {Kovner}}, \bibinfo {author} {\bibfnamefont {M.~A.}\ \bibnamefont
  {Shifman}},\ and\ \bibinfo {author} {\bibfnamefont {A.~V.}\ \bibnamefont
  {Smilga}},\ }\href {https://doi.org/10.1103/PhysRevD.56.7978} {\bibfield
  {journal} {\bibinfo  {journal} {Phys. Rev. D}\ }\textbf {\bibinfo {volume}
  {56}},\ \bibinfo {pages} {7978} (\bibinfo {year} {1997})},\ \Eprint
  {https://arxiv.org/abs/hep-th/9706089} {arXiv:hep-th/9706089} \BibitemShut
  {NoStop}%
\bibitem [{\citenamefont {Takahashi}\ \emph {et~al.}(2008)\citenamefont
  {Takahashi}, \citenamefont {Yanagida},\ and\ \citenamefont
  {Yonekura}}]{takahashi_gravitational_2008}%
  \BibitemOpen
  \bibfield  {author} {\bibinfo {author} {\bibfnamefont {F.}~\bibnamefont
  {Takahashi}}, \bibinfo {author} {\bibfnamefont {T.~T.}\ \bibnamefont
  {Yanagida}},\ and\ \bibinfo {author} {\bibfnamefont {K.}~\bibnamefont
  {Yonekura}},\ }\href
  {https://doi.org/https://doi.org/10.1016/j.physletb.2008.05.022} {\bibfield
  {journal} {\bibinfo  {journal} {Physics Letters B}\ }\textbf {\bibinfo
  {volume} {664}},\ \bibinfo {pages} {194} (\bibinfo {year}
  {2008})}\BibitemShut {NoStop}%
\bibitem [{\citenamefont {Dine}\ \emph {et~al.}(2010)\citenamefont {Dine},
  \citenamefont {Takahashi},\ and\ \citenamefont {Yanagida}}]{Dine:2010eb}%
  \BibitemOpen
  \bibfield  {author} {\bibinfo {author} {\bibfnamefont {M.}~\bibnamefont
  {Dine}}, \bibinfo {author} {\bibfnamefont {F.}~\bibnamefont {Takahashi}},\
  and\ \bibinfo {author} {\bibfnamefont {T.~T.}\ \bibnamefont {Yanagida}},\
  }\href {https://doi.org/10.1007/JHEP07(2010)003} {\bibfield  {journal}
  {\bibinfo  {journal} {JHEP}\ }\textbf {\bibinfo {volume} {07}},\ \bibinfo
  {pages} {003}},\ \Eprint {https://arxiv.org/abs/1005.3613} {arXiv:1005.3613
  [hep-th]} \BibitemShut {NoStop}%
\bibitem [{\citenamefont {Buttazzo}\ \emph {et~al.}(2013)\citenamefont
  {Buttazzo}, \citenamefont {Degrassi}, \citenamefont {Giardino}, \citenamefont
  {Giudice}, \citenamefont {Sala}, \citenamefont {Salvio},\ and\ \citenamefont
  {Strumia}}]{Buttazzo:2013uya}%
  \BibitemOpen
  \bibfield  {author} {\bibinfo {author} {\bibfnamefont {D.}~\bibnamefont
  {Buttazzo}}, \bibinfo {author} {\bibfnamefont {G.}~\bibnamefont {Degrassi}},
  \bibinfo {author} {\bibfnamefont {P.~P.}\ \bibnamefont {Giardino}}, \bibinfo
  {author} {\bibfnamefont {G.~F.}\ \bibnamefont {Giudice}}, \bibinfo {author}
  {\bibfnamefont {F.}~\bibnamefont {Sala}}, \bibinfo {author} {\bibfnamefont
  {A.}~\bibnamefont {Salvio}},\ and\ \bibinfo {author} {\bibfnamefont
  {A.}~\bibnamefont {Strumia}},\ }\href
  {https://doi.org/10.1007/JHEP12(2013)089} {\bibfield  {journal} {\bibinfo
  {journal} {JHEP}\ }\textbf {\bibinfo {volume} {12}},\ \bibinfo {pages}
  {089}},\ \Eprint {https://arxiv.org/abs/1307.3536} {arXiv:1307.3536 [hep-ph]}
  \BibitemShut {NoStop}%
\bibitem [{\citenamefont {Andreassen}\ \emph {et~al.}(2014)\citenamefont
  {Andreassen}, \citenamefont {Frost},\ and\ \citenamefont
  {Schwartz}}]{Andreassen:2014gha}%
  \BibitemOpen
  \bibfield  {author} {\bibinfo {author} {\bibfnamefont {A.}~\bibnamefont
  {Andreassen}}, \bibinfo {author} {\bibfnamefont {W.}~\bibnamefont {Frost}},\
  and\ \bibinfo {author} {\bibfnamefont {M.~D.}\ \bibnamefont {Schwartz}},\
  }\href {https://doi.org/10.1103/PhysRevLett.113.241801} {\bibfield  {journal}
  {\bibinfo  {journal} {Phys. Rev. Lett.}\ }\textbf {\bibinfo {volume} {113}},\
  \bibinfo {pages} {241801} (\bibinfo {year} {2014})},\ \Eprint
  {https://arxiv.org/abs/1408.0292} {arXiv:1408.0292 [hep-ph]} \BibitemShut
  {NoStop}%
\bibitem [{\citenamefont {Krajewski}\ \emph {et~al.}(2016)\citenamefont
  {Krajewski}, \citenamefont {Lalak}, \citenamefont {Lewicki},\ and\
  \citenamefont {Olszewski}}]{Krajewski:2016vbr}%
  \BibitemOpen
  \bibfield  {author} {\bibinfo {author} {\bibfnamefont {T.}~\bibnamefont
  {Krajewski}}, \bibinfo {author} {\bibfnamefont {Z.}~\bibnamefont {Lalak}},
  \bibinfo {author} {\bibfnamefont {M.}~\bibnamefont {Lewicki}},\ and\ \bibinfo
  {author} {\bibfnamefont {P.}~\bibnamefont {Olszewski}},\ }\href
  {https://doi.org/10.1088/1475-7516/2016/12/036} {\bibfield  {journal}
  {\bibinfo  {journal} {JCAP}\ }\textbf {\bibinfo {volume} {12}},\ \bibinfo
  {pages} {036}},\ \Eprint {https://arxiv.org/abs/1608.05719} {arXiv:1608.05719
  [astro-ph.CO]} \BibitemShut {NoStop}%
\bibitem [{\citenamefont {Vilenkin}(1985{\natexlab{a}})}]{Vilenkin:1984ib}%
  \BibitemOpen
  \bibfield  {author} {\bibinfo {author} {\bibfnamefont {A.}~\bibnamefont
  {Vilenkin}},\ }\href {https://doi.org/10.1016/0370-1573(85)90033-X}
  {\bibfield  {journal} {\bibinfo  {journal} {Phys. Rept.}\ }\textbf {\bibinfo
  {volume} {121}},\ \bibinfo {pages} {263} (\bibinfo {year}
  {1985}{\natexlab{a}})}\BibitemShut {NoStop}%
\bibitem [{\citenamefont {Hill}\ \emph {et~al.}(1989)\citenamefont {Hill},
  \citenamefont {Schramm},\ and\ \citenamefont {Fry}}]{Hill:1988vm}%
  \BibitemOpen
  \bibfield  {author} {\bibinfo {author} {\bibfnamefont {C.~T.}\ \bibnamefont
  {Hill}}, \bibinfo {author} {\bibfnamefont {D.~N.}\ \bibnamefont {Schramm}},\
  and\ \bibinfo {author} {\bibfnamefont {J.~N.}\ \bibnamefont {Fry}},\
  }\href@noop {} {\bibfield  {journal} {\bibinfo  {journal} {Comments Nucl.
  Part. Phys.}\ }\textbf {\bibinfo {volume} {19}},\ \bibinfo {pages} {25}
  (\bibinfo {year} {1989})}\BibitemShut {NoStop}%
\bibitem [{\citenamefont {Takahashi}\ and\ \citenamefont
  {Yin}(2021)}]{Takahashi:2020tqv}%
  \BibitemOpen
  \bibfield  {author} {\bibinfo {author} {\bibfnamefont {F.}~\bibnamefont
  {Takahashi}}\ and\ \bibinfo {author} {\bibfnamefont {W.}~\bibnamefont
  {Yin}},\ }\href {https://doi.org/10.1088/1475-7516/2021/04/007} {\bibfield
  {journal} {\bibinfo  {journal} {JCAP}\ }\textbf {\bibinfo {volume} {04}},\
  \bibinfo {pages} {007}},\ \Eprint {https://arxiv.org/abs/2012.11576}
  {arXiv:2012.11576 [hep-ph]} \BibitemShut {NoStop}%
\bibitem [{\citenamefont {Gonzalez}\ \emph {et~al.}(2022)\citenamefont
  {Gonzalez}, \citenamefont {Kitajima}, \citenamefont {Takahashi},\ and\
  \citenamefont {Yin}}]{Gonzalez:2022mcx}%
  \BibitemOpen
  \bibfield  {author} {\bibinfo {author} {\bibfnamefont {D.}~\bibnamefont
  {Gonzalez}}, \bibinfo {author} {\bibfnamefont {N.}~\bibnamefont {Kitajima}},
  \bibinfo {author} {\bibfnamefont {F.}~\bibnamefont {Takahashi}},\ and\
  \bibinfo {author} {\bibfnamefont {W.}~\bibnamefont {Yin}},\ }\href@noop {} {\
   (\bibinfo {year} {2022})},\ \Eprint {https://arxiv.org/abs/2211.06849}
  {arXiv:2211.06849 [hep-ph]} \BibitemShut {NoStop}%
\bibitem [{\citenamefont {Hiramatsu}\ \emph {et~al.}(2014)\citenamefont
  {Hiramatsu}, \citenamefont {Kawasaki},\ and\ \citenamefont
  {Saikawa}}]{Hiramatsu:2013qaa}%
  \BibitemOpen
  \bibfield  {author} {\bibinfo {author} {\bibfnamefont {T.}~\bibnamefont
  {Hiramatsu}}, \bibinfo {author} {\bibfnamefont {M.}~\bibnamefont
  {Kawasaki}},\ and\ \bibinfo {author} {\bibfnamefont {K.}~\bibnamefont
  {Saikawa}},\ }\href {https://doi.org/10.1088/1475-7516/2014/02/031}
  {\bibfield  {journal} {\bibinfo  {journal} {JCAP}\ }\textbf {\bibinfo
  {volume} {02}},\ \bibinfo {pages} {031}},\ \Eprint
  {https://arxiv.org/abs/1309.5001} {arXiv:1309.5001 [astro-ph.CO]}
  \BibitemShut {NoStop}%
\bibitem [{\citenamefont {Saikawa}(2017{\natexlab{a}})}]{Saikawa:2017hiv}%
  \BibitemOpen
  \bibfield  {author} {\bibinfo {author} {\bibfnamefont {K.}~\bibnamefont
  {Saikawa}},\ }\href {https://doi.org/10.3390/universe3020040} {\bibfield
  {journal} {\bibinfo  {journal} {Universe}\ }\textbf {\bibinfo {volume} {3}},\
  \bibinfo {pages} {40} (\bibinfo {year} {2017}{\natexlab{a}})},\ \Eprint
  {https://arxiv.org/abs/1703.02576} {arXiv:1703.02576 [hep-ph]} \BibitemShut
  {NoStop}%
\bibitem [{\citenamefont {Hiramatsu}\ \emph
  {et~al.}(2013{\natexlab{b}})\citenamefont {Hiramatsu}, \citenamefont
  {Kawasaki}, \citenamefont {Saikawa},\ and\ \citenamefont
  {Sekiguchi}}]{hiramatsu_axion_2013}%
  \BibitemOpen
  \bibfield  {author} {\bibinfo {author} {\bibfnamefont {T.}~\bibnamefont
  {Hiramatsu}}, \bibinfo {author} {\bibfnamefont {M.}~\bibnamefont {Kawasaki}},
  \bibinfo {author} {\bibfnamefont {K.}~\bibnamefont {Saikawa}},\ and\ \bibinfo
  {author} {\bibfnamefont {T.}~\bibnamefont {Sekiguchi}},\ }\href
  {https://doi.org/10.1088/1475-7516/2013/01/001} {\bibfield  {journal}
  {\bibinfo  {journal} {Journal of Cosmology and Astroparticle Physics}\
  }\textbf {\bibinfo {volume} {2013}}\bibfield  {number} {\bibinfo  {number} {
  (01)},\ \bibinfo {pages} {001}},\ }\bibinfo {note} {arXiv:1207.3166
  [astro-ph, physics:hep-ph]}\BibitemShut {NoStop}%
\bibitem [{\citenamefont {Wei}\ and\ \citenamefont
  {Jiang}(2022)}]{Wei:2022poh}%
  \BibitemOpen
  \bibfield  {author} {\bibinfo {author} {\bibfnamefont {D.}~\bibnamefont
  {Wei}}\ and\ \bibinfo {author} {\bibfnamefont {Y.}~\bibnamefont {Jiang}},\
  }\href@noop {} {\  (\bibinfo {year} {2022})},\ \Eprint
  {https://arxiv.org/abs/2208.07186} {arXiv:2208.07186 [hep-ph]} \BibitemShut
  {NoStop}%
\bibitem [{\citenamefont {Sakharov}\ \emph {et~al.}(2021)\citenamefont
  {Sakharov}, \citenamefont {Eroshenko},\ and\ \citenamefont
  {Rubin}}]{PhysRevD.104.043005}%
  \BibitemOpen
  \bibfield  {author} {\bibinfo {author} {\bibfnamefont {A.~S.}\ \bibnamefont
  {Sakharov}}, \bibinfo {author} {\bibfnamefont {Y.~N.}\ \bibnamefont
  {Eroshenko}},\ and\ \bibinfo {author} {\bibfnamefont {S.~G.}\ \bibnamefont
  {Rubin}},\ }\href {https://doi.org/10.1103/PhysRevD.104.043005} {\bibfield
  {journal} {\bibinfo  {journal} {Phys. Rev. D}\ }\textbf {\bibinfo {volume}
  {104}},\ \bibinfo {pages} {043005} (\bibinfo {year} {2021})}\BibitemShut
  {NoStop}%
\bibitem [{\citenamefont {Liu}\ \emph {et~al.}(2021)\citenamefont {Liu},
  \citenamefont {Cai},\ and\ \citenamefont {Guo}}]{Liu:2020mru}%
  \BibitemOpen
  \bibfield  {author} {\bibinfo {author} {\bibfnamefont {J.}~\bibnamefont
  {Liu}}, \bibinfo {author} {\bibfnamefont {R.-G.}\ \bibnamefont {Cai}},\ and\
  \bibinfo {author} {\bibfnamefont {Z.-K.}\ \bibnamefont {Guo}},\ }\href
  {https://doi.org/10.1103/PhysRevLett.126.141303} {\bibfield  {journal}
  {\bibinfo  {journal} {Phys. Rev. Lett.}\ }\textbf {\bibinfo {volume} {126}},\
  \bibinfo {pages} {141303} (\bibinfo {year} {2021})},\ \Eprint
  {https://arxiv.org/abs/2010.03225} {arXiv:2010.03225 [astro-ph.CO]}
  \BibitemShut {NoStop}%
\bibitem [{\citenamefont {Blasi}\ and\ \citenamefont
  {Mariotti}(2022)}]{Blasi:2022woz}%
  \BibitemOpen
  \bibfield  {author} {\bibinfo {author} {\bibfnamefont {S.}~\bibnamefont
  {Blasi}}\ and\ \bibinfo {author} {\bibfnamefont {A.}~\bibnamefont
  {Mariotti}},\ }\href {https://doi.org/10.1103/PhysRevLett.129.261303}
  {\bibfield  {journal} {\bibinfo  {journal} {Phys. Rev. Lett.}\ }\textbf
  {\bibinfo {volume} {129}},\ \bibinfo {pages} {261303} (\bibinfo {year}
  {2022})},\ \Eprint {https://arxiv.org/abs/2203.16450} {arXiv:2203.16450
  [hep-ph]} \BibitemShut {NoStop}%
\bibitem [{\citenamefont
  {Vilenkin}(1985{\natexlab{b}})}]{vilenkin_cosmic_1985}%
  \BibitemOpen
  \bibfield  {author} {\bibinfo {author} {\bibfnamefont {A.}~\bibnamefont
  {Vilenkin}},\ }\href
  {https://doi.org/https://doi.org/10.1016/0370-1573(85)90033-X} {\bibfield
  {journal} {\bibinfo  {journal} {Physics Reports}\ }\textbf {\bibinfo {volume}
  {121}},\ \bibinfo {pages} {263} (\bibinfo {year}
  {1985}{\natexlab{b}})}\BibitemShut {NoStop}%
\bibitem [{\citenamefont {Saikawa}(2017{\natexlab{b}})}]{saikawa_review_2017}%
  \BibitemOpen
  \bibfield  {author} {\bibinfo {author} {\bibfnamefont {K.}~\bibnamefont
  {Saikawa}},\ }\href {https://doi.org/10.3390/universe3020040} {\bibfield
  {journal} {\bibinfo  {journal} {Universe}\ }\textbf {\bibinfo {volume} {3}},\
  \bibinfo {pages} {40} (\bibinfo {year} {2017}{\natexlab{b}})},\ \bibinfo
  {note} {arXiv:1703.02576 [astro-ph, physics:gr-qc,
  physics:hep-ph]}\BibitemShut {NoStop}%
\bibitem [{\citenamefont {{Press}}\ \emph {et~al.}(1989)\citenamefont
  {{Press}}, \citenamefont {{Ryden}},\ and\ \citenamefont
  {{Spergel}}}]{1989ApJ...347..590P}%
  \BibitemOpen
  \bibfield  {author} {\bibinfo {author} {\bibfnamefont {W.~H.}\ \bibnamefont
  {{Press}}}, \bibinfo {author} {\bibfnamefont {B.~S.}\ \bibnamefont
  {{Ryden}}},\ and\ \bibinfo {author} {\bibfnamefont {D.~N.}\ \bibnamefont
  {{Spergel}}},\ }\href {https://doi.org/10.1086/168151} {\bibfield  {journal}
  {\bibinfo  {journal} {\apj}\ }\textbf {\bibinfo {volume} {347}},\ \bibinfo
  {pages} {590} (\bibinfo {year} {1989})}\BibitemShut {NoStop}%
\bibitem [{\citenamefont {Hindmarsh}(1996)}]{hindmarsh_analytic_1996}%
  \BibitemOpen
  \bibfield  {author} {\bibinfo {author} {\bibfnamefont {M.}~\bibnamefont
  {Hindmarsh}},\ }\href {https://doi.org/10.1103/PhysRevLett.77.4495}
  {\bibfield  {journal} {\bibinfo  {journal} {Physical Review Letters}\
  }\textbf {\bibinfo {volume} {77}},\ \bibinfo {pages} {4495} (\bibinfo {year}
  {1996})},\ \bibinfo {note} {arXiv:hep-ph/9605332}\BibitemShut {NoStop}%
\bibitem [{\citenamefont {Martins}\ \emph {et~al.}(2016)\citenamefont
  {Martins}, \citenamefont {Rybak}, \citenamefont {Avgoustidis},\ and\
  \citenamefont {Shellard}}]{martins_extending_2016}%
  \BibitemOpen
  \bibfield  {author} {\bibinfo {author} {\bibfnamefont {C.~J. A.~P.}\
  \bibnamefont {Martins}}, \bibinfo {author} {\bibfnamefont {I.~Y.}\
  \bibnamefont {Rybak}}, \bibinfo {author} {\bibfnamefont {A.}~\bibnamefont
  {Avgoustidis}},\ and\ \bibinfo {author} {\bibfnamefont {E.~P.~S.}\
  \bibnamefont {Shellard}},\ }\href
  {https://doi.org/10.1103/PhysRevD.93.043534} {\bibfield  {journal} {\bibinfo
  {journal} {Physical Review D}\ }\textbf {\bibinfo {volume} {93}},\ \bibinfo
  {pages} {043534} (\bibinfo {year} {2016})},\ \bibinfo {note}
  {arXiv:1602.01322 [astro-ph, physics:gr-qc, physics:hep-ph,
  physics:hep-th]}\BibitemShut {NoStop}%
\bibitem [{\citenamefont {Leite}\ and\ \citenamefont
  {Martins}(2011)}]{leite_scaling_2011}%
  \BibitemOpen
  \bibfield  {author} {\bibinfo {author} {\bibfnamefont {A.~M.~M.}\
  \bibnamefont {Leite}}\ and\ \bibinfo {author} {\bibfnamefont {C.~J. A.~P.}\
  \bibnamefont {Martins}},\ }\href {https://doi.org/10.1103/PhysRevD.84.103523}
  {\bibfield  {journal} {\bibinfo  {journal} {Physical Review D}\ }\textbf
  {\bibinfo {volume} {84}},\ \bibinfo {pages} {103523} (\bibinfo {year}
  {2011})},\ \bibinfo {note} {arXiv:1110.3486 [astro-ph, physics:hep-ph,
  physics:hep-th]}\BibitemShut {NoStop}%
\bibitem [{\citenamefont {Leite}\ \emph {et~al.}(2013)\citenamefont {Leite},
  \citenamefont {Martins},\ and\ \citenamefont
  {Shellard}}]{leite_accurate_2013}%
  \BibitemOpen
  \bibfield  {author} {\bibinfo {author} {\bibfnamefont {A.~M.~M.}\
  \bibnamefont {Leite}}, \bibinfo {author} {\bibfnamefont {C.~J. A.~P.}\
  \bibnamefont {Martins}},\ and\ \bibinfo {author} {\bibfnamefont {E.~P.~S.}\
  \bibnamefont {Shellard}},\ }\href
  {https://doi.org/10.1016/j.physletb.2012.11.070} {\bibfield  {journal}
  {\bibinfo  {journal} {Physics Letters B}\ }\textbf {\bibinfo {volume}
  {718}},\ \bibinfo {pages} {740} (\bibinfo {year} {2013})},\ \bibinfo {note}
  {arXiv:1206.6043 [hep-ph, physics:hep-th]}\BibitemShut {NoStop}%
\bibitem [{\citenamefont {Collaboration}\ \emph {et~al.}(2020)\citenamefont
  {Collaboration}, \citenamefont {Akrami},\ and\ \citenamefont {et~\quad
  al}}]{planck_collaboration_planck_2020}%
  \BibitemOpen
  \bibfield  {author} {\bibinfo {author} {\bibfnamefont {P.}~\bibnamefont
  {Collaboration}}, \bibinfo {author} {\bibfnamefont {Y.}~\bibnamefont
  {Akrami}},\ and\ \bibinfo {author} {\bibnamefont {et~\quad al}},\ }\href
  {https://doi.org/10.1051/0004-6361/201833880} {\bibfield  {journal} {\bibinfo
   {journal} {Astronomy \& Astrophysics}\ }\textbf {\bibinfo {volume} {641}},\
  \bibinfo {pages} {A1} (\bibinfo {year} {2020})},\ \bibinfo {note} {arXiv:
  1807.06205}\BibitemShut {NoStop}%
\bibitem [{\citenamefont {Jiang}\ and\ \citenamefont
  {Huang}(2022)}]{Jiang:2022svq}%
  \BibitemOpen
  \bibfield  {author} {\bibinfo {author} {\bibfnamefont {Y.}~\bibnamefont
  {Jiang}}\ and\ \bibinfo {author} {\bibfnamefont {Q.-G.}\ \bibnamefont
  {Huang}},\ }\href {https://doi.org/10.1103/PhysRevD.106.103036} {\bibfield
  {journal} {\bibinfo  {journal} {Phys. Rev. D}\ }\textbf {\bibinfo {volume}
  {106}},\ \bibinfo {pages} {103036} (\bibinfo {year} {2022})},\ \Eprint
  {https://arxiv.org/abs/2208.00697} {arXiv:2208.00697 [astro-ph.CO]}
  \BibitemShut {NoStop}%
\bibitem [{\citenamefont {Bian}\ \emph {et~al.}(2022)\citenamefont {Bian},
  \citenamefont {Ge}, \citenamefont {Li}, \citenamefont {Shu},\ and\
  \citenamefont {Zong}}]{Bian:2022qbh}%
  \BibitemOpen
  \bibfield  {author} {\bibinfo {author} {\bibfnamefont {L.}~\bibnamefont
  {Bian}}, \bibinfo {author} {\bibfnamefont {S.}~\bibnamefont {Ge}}, \bibinfo
  {author} {\bibfnamefont {C.}~\bibnamefont {Li}}, \bibinfo {author}
  {\bibfnamefont {J.}~\bibnamefont {Shu}},\ and\ \bibinfo {author}
  {\bibfnamefont {J.}~\bibnamefont {Zong}},\ }\href@noop {} {\  (\bibinfo
  {year} {2022})},\ \Eprint {https://arxiv.org/abs/2212.07871}
  {arXiv:2212.07871 [hep-ph]} \BibitemShut {NoStop}%
\bibitem [{\citenamefont {Ferreira}\ \emph {et~al.}(2022)\citenamefont
  {Ferreira}, \citenamefont {Notari}, \citenamefont {Pujolas},\ and\
  \citenamefont {Rompineve}}]{Ferreira:2022zzo}%
  \BibitemOpen
  \bibfield  {author} {\bibinfo {author} {\bibfnamefont {R.~Z.}\ \bibnamefont
  {Ferreira}}, \bibinfo {author} {\bibfnamefont {A.}~\bibnamefont {Notari}},
  \bibinfo {author} {\bibfnamefont {O.}~\bibnamefont {Pujolas}},\ and\ \bibinfo
  {author} {\bibfnamefont {F.}~\bibnamefont {Rompineve}},\ }\href@noop {} {\
  (\bibinfo {year} {2022})},\ \Eprint {https://arxiv.org/abs/2204.04228}
  {arXiv:2204.04228 [astro-ph.CO]} \BibitemShut {NoStop}%
\bibitem [{\citenamefont {Ramberg}\ \emph {et~al.}(2022)\citenamefont
  {Ramberg}, \citenamefont {Ratzinger},\ and\ \citenamefont
  {Schwaller}}]{Ramberg:2022irf}%
  \BibitemOpen
  \bibfield  {author} {\bibinfo {author} {\bibfnamefont {N.}~\bibnamefont
  {Ramberg}}, \bibinfo {author} {\bibfnamefont {W.}~\bibnamefont {Ratzinger}},\
  and\ \bibinfo {author} {\bibfnamefont {P.}~\bibnamefont {Schwaller}},\
  }\href@noop {} {\  (\bibinfo {year} {2022})},\ \Eprint
  {https://arxiv.org/abs/2209.14313} {arXiv:2209.14313 [hep-ph]} \BibitemShut
  {NoStop}%
\bibitem [{\citenamefont {Basu}\ \emph {et~al.}(1991)\citenamefont {Basu},
  \citenamefont {Guth},\ and\ \citenamefont {Vilenkin}}]{basu_quantum_1991}%
  \BibitemOpen
  \bibfield  {author} {\bibinfo {author} {\bibfnamefont {R.}~\bibnamefont
  {Basu}}, \bibinfo {author} {\bibfnamefont {A.~H.}\ \bibnamefont {Guth}},\
  and\ \bibinfo {author} {\bibfnamefont {A.}~\bibnamefont {Vilenkin}},\ }\href
  {https://doi.org/10.1103/PhysRevD.44.340} {\bibfield  {journal} {\bibinfo
  {journal} {Physical Review D}\ }\textbf {\bibinfo {volume} {44}},\ \bibinfo
  {pages} {340} (\bibinfo {year} {1991})}\BibitemShut {NoStop}%
\bibitem [{\citenamefont {Liu}\ \emph {et~al.}(2020{\natexlab{a}})\citenamefont
  {Liu}, \citenamefont {Guo},\ and\ \citenamefont {Cai}}]{liu_primordial_2020}%
  \BibitemOpen
  \bibfield  {author} {\bibinfo {author} {\bibfnamefont {J.}~\bibnamefont
  {Liu}}, \bibinfo {author} {\bibfnamefont {Z.-K.}\ \bibnamefont {Guo}},\ and\
  \bibinfo {author} {\bibfnamefont {R.-G.}\ \bibnamefont {Cai}},\ }\href
  {https://doi.org/10.1103/PhysRevD.101.023513} {\bibfield  {journal} {\bibinfo
   {journal} {Physical Review D}\ }\textbf {\bibinfo {volume} {101}},\ \bibinfo
  {pages} {023513} (\bibinfo {year} {2020}{\natexlab{a}})},\ \bibinfo {note}
  {arXiv:1908.02662 [astro-ph, physics:gr-qc, physics:hep-th]}\BibitemShut
  {NoStop}%
\bibitem [{\citenamefont {Vachaspati}\ \emph {et~al.}(1984)\citenamefont
  {Vachaspati}, \citenamefont {Everett},\ and\ \citenamefont
  {Vilenkin}}]{Vachaspati:1984yi}%
  \BibitemOpen
  \bibfield  {author} {\bibinfo {author} {\bibfnamefont {T.}~\bibnamefont
  {Vachaspati}}, \bibinfo {author} {\bibfnamefont {A.~E.}\ \bibnamefont
  {Everett}},\ and\ \bibinfo {author} {\bibfnamefont {A.}~\bibnamefont
  {Vilenkin}},\ }\href {https://doi.org/10.1103/PhysRevD.30.2046} {\bibfield
  {journal} {\bibinfo  {journal} {Phys. Rev. D}\ }\textbf {\bibinfo {volume}
  {30}},\ \bibinfo {pages} {2046} (\bibinfo {year} {1984})}\BibitemShut
  {NoStop}%
\bibitem [{\citenamefont {Pujolas}\ and\ \citenamefont
  {Zahariade}(2022)}]{Pujolas:2022qvs}%
  \BibitemOpen
  \bibfield  {author} {\bibinfo {author} {\bibfnamefont {O.}~\bibnamefont
  {Pujolas}}\ and\ \bibinfo {author} {\bibfnamefont {G.}~\bibnamefont
  {Zahariade}},\ }\href@noop {} {\  (\bibinfo {year} {2022})},\ \Eprint
  {https://arxiv.org/abs/2212.11204} {arXiv:2212.11204 [hep-th]} \BibitemShut
  {NoStop}%
\bibitem [{\citenamefont {Blasi}\ \emph {et~al.}(2022)\citenamefont {Blasi},
  \citenamefont {Mariotti}, \citenamefont {Rase}, \citenamefont {Sevrin},\ and\
  \citenamefont {Turbang}}]{Blasi:2022ayo}%
  \BibitemOpen
  \bibfield  {author} {\bibinfo {author} {\bibfnamefont {S.}~\bibnamefont
  {Blasi}}, \bibinfo {author} {\bibfnamefont {A.}~\bibnamefont {Mariotti}},
  \bibinfo {author} {\bibfnamefont {A.}~\bibnamefont {Rase}}, \bibinfo {author}
  {\bibfnamefont {A.}~\bibnamefont {Sevrin}},\ and\ \bibinfo {author}
  {\bibfnamefont {K.}~\bibnamefont {Turbang}},\ }\href@noop {} {\  (\bibinfo
  {year} {2022})},\ \Eprint {https://arxiv.org/abs/2210.14246}
  {arXiv:2210.14246 [hep-ph]} \BibitemShut {NoStop}%
\bibitem [{\citenamefont {Tanahashi}\ and\ \citenamefont
  {Yoo}(2015)}]{Tanahashi:2014sma}%
  \BibitemOpen
  \bibfield  {author} {\bibinfo {author} {\bibfnamefont {N.}~\bibnamefont
  {Tanahashi}}\ and\ \bibinfo {author} {\bibfnamefont {C.-M.}\ \bibnamefont
  {Yoo}},\ }\href {https://doi.org/10.1088/0264-9381/32/15/155003} {\bibfield
  {journal} {\bibinfo  {journal} {Class. Quant. Grav.}\ }\textbf {\bibinfo
  {volume} {32}},\ \bibinfo {pages} {155003} (\bibinfo {year} {2015})},\
  \Eprint {https://arxiv.org/abs/1411.7479} {arXiv:1411.7479 [gr-qc]}
  \BibitemShut {NoStop}%
\bibitem [{\citenamefont {Garriga}\ \emph {et~al.}(2016)\citenamefont
  {Garriga}, \citenamefont {Vilenkin},\ and\ \citenamefont
  {Zhang}}]{Garriga:2015fdk}%
  \BibitemOpen
  \bibfield  {author} {\bibinfo {author} {\bibfnamefont {J.}~\bibnamefont
  {Garriga}}, \bibinfo {author} {\bibfnamefont {A.}~\bibnamefont {Vilenkin}},\
  and\ \bibinfo {author} {\bibfnamefont {J.}~\bibnamefont {Zhang}},\ }\href
  {https://doi.org/10.1088/1475-7516/2016/02/064} {\bibfield  {journal}
  {\bibinfo  {journal} {JCAP}\ }\textbf {\bibinfo {volume} {02}},\ \bibinfo
  {pages} {064}},\ \Eprint {https://arxiv.org/abs/1512.01819} {arXiv:1512.01819
  [hep-th]} \BibitemShut {NoStop}%
\bibitem [{\citenamefont {Deng}\ \emph
  {et~al.}(2017{\natexlab{a}})\citenamefont {Deng}, \citenamefont {Garriga},\
  and\ \citenamefont {Vilenkin}}]{Deng:2016vzb}%
  \BibitemOpen
  \bibfield  {author} {\bibinfo {author} {\bibfnamefont {H.}~\bibnamefont
  {Deng}}, \bibinfo {author} {\bibfnamefont {J.}~\bibnamefont {Garriga}},\ and\
  \bibinfo {author} {\bibfnamefont {A.}~\bibnamefont {Vilenkin}},\ }\href
  {https://doi.org/10.1088/1475-7516/2017/04/050} {\bibfield  {journal}
  {\bibinfo  {journal} {JCAP}\ }\textbf {\bibinfo {volume} {04}},\ \bibinfo
  {pages} {050}},\ \Eprint {https://arxiv.org/abs/1612.03753} {arXiv:1612.03753
  [gr-qc]} \BibitemShut {NoStop}%
\bibitem [{\citenamefont {Liu}\ \emph {et~al.}(2020{\natexlab{b}})\citenamefont
  {Liu}, \citenamefont {Guo},\ and\ \citenamefont {Cai}}]{Liu:2019lul}%
  \BibitemOpen
  \bibfield  {author} {\bibinfo {author} {\bibfnamefont {J.}~\bibnamefont
  {Liu}}, \bibinfo {author} {\bibfnamefont {Z.-K.}\ \bibnamefont {Guo}},\ and\
  \bibinfo {author} {\bibfnamefont {R.-G.}\ \bibnamefont {Cai}},\ }\href
  {https://doi.org/10.1103/PhysRevD.101.023513} {\bibfield  {journal} {\bibinfo
   {journal} {Phys. Rev. D}\ }\textbf {\bibinfo {volume} {101}},\ \bibinfo
  {pages} {023513} (\bibinfo {year} {2020}{\natexlab{b}})},\ \Eprint
  {https://arxiv.org/abs/1908.02662} {arXiv:1908.02662 [astro-ph.CO]}
  \BibitemShut {NoStop}%
\bibitem [{\citenamefont {Ge}(2020)}]{Ge:2019ihf}%
  \BibitemOpen
  \bibfield  {author} {\bibinfo {author} {\bibfnamefont {S.}~\bibnamefont
  {Ge}},\ }\href {https://doi.org/10.1016/j.dark.2019.100440} {\bibfield
  {journal} {\bibinfo  {journal} {Phys. Dark Univ.}\ }\textbf {\bibinfo
  {volume} {27}},\ \bibinfo {pages} {100440} (\bibinfo {year} {2020})},\
  \Eprint {https://arxiv.org/abs/1905.12182} {arXiv:1905.12182 [hep-ph]}
  \BibitemShut {NoStop}%
\bibitem [{\citenamefont {Garriga}(1994)}]{garriga_nucleation_1994}%
  \BibitemOpen
  \bibfield  {author} {\bibinfo {author} {\bibfnamefont {J.}~\bibnamefont
  {Garriga}},\ }\href {https://doi.org/10.1103/PhysRevD.49.6327} {\bibfield
  {journal} {\bibinfo  {journal} {Physical Review D}\ }\textbf {\bibinfo
  {volume} {49}},\ \bibinfo {pages} {6327} (\bibinfo {year} {1994})},\ \bibinfo
  {note} {arXiv:hep-ph/9308280}\BibitemShut {NoStop}%
\bibitem [{\citenamefont {Bardeen}\ \emph {et~al.}(1986)\citenamefont
  {Bardeen}, \citenamefont {Bond}, \citenamefont {Kaiser},\ and\ \citenamefont
  {Szalay}}]{Bardeen:1985tr}%
  \BibitemOpen
  \bibfield  {author} {\bibinfo {author} {\bibfnamefont {J.~M.}\ \bibnamefont
  {Bardeen}}, \bibinfo {author} {\bibfnamefont {J.~R.}\ \bibnamefont {Bond}},
  \bibinfo {author} {\bibfnamefont {N.}~\bibnamefont {Kaiser}},\ and\ \bibinfo
  {author} {\bibfnamefont {A.~S.}\ \bibnamefont {Szalay}},\ }\href
  {https://doi.org/10.1086/164143} {\bibfield  {journal} {\bibinfo  {journal}
  {Astrophys. J.}\ }\textbf {\bibinfo {volume} {304}},\ \bibinfo {pages} {15}
  (\bibinfo {year} {1986})}\BibitemShut {NoStop}%
\bibitem [{\citenamefont {Deng}\ \emph
  {et~al.}(2017{\natexlab{b}})\citenamefont {Deng}, \citenamefont {Garriga},\
  and\ \citenamefont {Vilenkin}}]{deng_primordial_2017}%
  \BibitemOpen
  \bibfield  {author} {\bibinfo {author} {\bibfnamefont {H.}~\bibnamefont
  {Deng}}, \bibinfo {author} {\bibfnamefont {J.}~\bibnamefont {Garriga}},\ and\
  \bibinfo {author} {\bibfnamefont {A.}~\bibnamefont {Vilenkin}},\ }\href
  {https://doi.org/10.1088/1475-7516/2017/04/050} {\bibfield  {journal}
  {\bibinfo  {journal} {Journal of Cosmology and Astroparticle Physics}\
  }\textbf {\bibinfo {volume} {2017}}\bibfield  {number} {\bibinfo  {number} {
  (04)},\ \bibinfo {pages} {050}},\ }\bibinfo {note} {arXiv:1612.03753 [gr-qc,
  physics:hep-th]}\BibitemShut {NoStop}%
\bibitem [{\citenamefont {Espinosa}\ \emph {et~al.}(2018)\citenamefont
  {Espinosa}, \citenamefont {Racco},\ and\ \citenamefont
  {Riotto}}]{Espinosa:2018eve}%
  \BibitemOpen
  \bibfield  {author} {\bibinfo {author} {\bibfnamefont {J.~R.}\ \bibnamefont
  {Espinosa}}, \bibinfo {author} {\bibfnamefont {D.}~\bibnamefont {Racco}},\
  and\ \bibinfo {author} {\bibfnamefont {A.}~\bibnamefont {Riotto}},\ }\href
  {https://doi.org/10.1088/1475-7516/2018/09/012} {\bibfield  {journal}
  {\bibinfo  {journal} {JCAP}\ }\textbf {\bibinfo {volume} {09}},\ \bibinfo
  {pages} {012}},\ \Eprint {https://arxiv.org/abs/1804.07732} {arXiv:1804.07732
  [hep-ph]} \BibitemShut {NoStop}%
\bibitem [{\citenamefont {Kohri}\ and\ \citenamefont
  {Terada}(2018)}]{Kohri:2018awv}%
  \BibitemOpen
  \bibfield  {author} {\bibinfo {author} {\bibfnamefont {K.}~\bibnamefont
  {Kohri}}\ and\ \bibinfo {author} {\bibfnamefont {T.}~\bibnamefont {Terada}},\
  }\href {https://doi.org/10.1103/PhysRevD.97.123532} {\bibfield  {journal}
  {\bibinfo  {journal} {Phys. Rev. D}\ }\textbf {\bibinfo {volume} {97}},\
  \bibinfo {pages} {123532} (\bibinfo {year} {2018})},\ \Eprint
  {https://arxiv.org/abs/1804.08577} {arXiv:1804.08577 [gr-qc]} \BibitemShut
  {NoStop}%
\bibitem [{\citenamefont {Emami}\ and\ \citenamefont
  {Smoot}(2018)}]{Emami:2017fiy}%
  \BibitemOpen
  \bibfield  {author} {\bibinfo {author} {\bibfnamefont {R.}~\bibnamefont
  {Emami}}\ and\ \bibinfo {author} {\bibfnamefont {G.}~\bibnamefont {Smoot}},\
  }\href {https://doi.org/10.1088/1475-7516/2018/01/007} {\bibfield  {journal}
  {\bibinfo  {journal} {JCAP}\ }\textbf {\bibinfo {volume} {01}},\ \bibinfo
  {pages} {007}},\ \Eprint {https://arxiv.org/abs/1705.09924} {arXiv:1705.09924
  [astro-ph.CO]} \BibitemShut {NoStop}%
\bibitem [{\citenamefont {Gow}\ \emph {et~al.}(2021)\citenamefont {Gow},
  \citenamefont {Byrnes}, \citenamefont {Cole},\ and\ \citenamefont
  {Young}}]{Gow:2020bzo}%
  \BibitemOpen
  \bibfield  {author} {\bibinfo {author} {\bibfnamefont {A.~D.}\ \bibnamefont
  {Gow}}, \bibinfo {author} {\bibfnamefont {C.~T.}\ \bibnamefont {Byrnes}},
  \bibinfo {author} {\bibfnamefont {P.~S.}\ \bibnamefont {Cole}},\ and\
  \bibinfo {author} {\bibfnamefont {S.}~\bibnamefont {Young}},\ }\href
  {https://doi.org/10.1088/1475-7516/2021/02/002} {\bibfield  {journal}
  {\bibinfo  {journal} {JCAP}\ }\textbf {\bibinfo {volume} {02}},\ \bibinfo
  {pages} {002}},\ \Eprint {https://arxiv.org/abs/2008.03289} {arXiv:2008.03289
  [astro-ph.CO]} \BibitemShut {NoStop}%
\bibitem [{\citenamefont {Lentati}\ \emph {et~al.}(2015)\citenamefont {Lentati}
  \emph {et~al.}}]{Lentati:2015qwp}%
  \BibitemOpen
  \bibfield  {author} {\bibinfo {author} {\bibfnamefont {L.}~\bibnamefont
  {Lentati}} \emph {et~al.},\ }\href {https://doi.org/10.1093/mnras/stv1538}
  {\bibfield  {journal} {\bibinfo  {journal} {Mon. Not. Roy. Astron. Soc.}\
  }\textbf {\bibinfo {volume} {453}},\ \bibinfo {pages} {2576} (\bibinfo {year}
  {2015})},\ \Eprint {https://arxiv.org/abs/1504.03692} {arXiv:1504.03692
  [astro-ph.CO]} \BibitemShut {NoStop}%
\bibitem [{\citenamefont {Shannon}\ \emph {et~al.}(2015)\citenamefont {Shannon}
  \emph {et~al.}}]{Shannon:2015ect}%
  \BibitemOpen
  \bibfield  {author} {\bibinfo {author} {\bibfnamefont {R.~M.}\ \bibnamefont
  {Shannon}} \emph {et~al.},\ }\href {https://doi.org/10.1126/science.aab1910}
  {\bibfield  {journal} {\bibinfo  {journal} {Science}\ }\textbf {\bibinfo
  {volume} {349}},\ \bibinfo {pages} {1522} (\bibinfo {year} {2015})},\ \Eprint
  {https://arxiv.org/abs/1509.07320} {arXiv:1509.07320 [astro-ph.CO]}
  \BibitemShut {NoStop}%
\bibitem [{\citenamefont {Arzoumanian}\ \emph {et~al.}(2018)\citenamefont
  {Arzoumanian} \emph {et~al.}}]{Arzoumanian:2018saf}%
  \BibitemOpen
  \bibfield  {author} {\bibinfo {author} {\bibfnamefont {Z.}~\bibnamefont
  {Arzoumanian}} \emph {et~al.} (\bibinfo {collaboration} {NANOGRAV}),\ }\href
  {https://doi.org/10.3847/1538-4357/aabd3b} {\bibfield  {journal} {\bibinfo
  {journal} {Astrophys. J.}\ }\textbf {\bibinfo {volume} {859}},\ \bibinfo
  {pages} {47} (\bibinfo {year} {2018})},\ \Eprint
  {https://arxiv.org/abs/1801.02617} {arXiv:1801.02617 [astro-ph.HE]}
  \BibitemShut {NoStop}%
\bibitem [{\citenamefont {Arzoumanian}\ \emph {et~al.}(2020)\citenamefont
  {Arzoumanian} \emph {et~al.}}]{NANOGrav:2020bcs}%
  \BibitemOpen
  \bibfield  {author} {\bibinfo {author} {\bibfnamefont {Z.}~\bibnamefont
  {Arzoumanian}} \emph {et~al.} (\bibinfo {collaboration} {NANOGrav}),\ }\href
  {https://doi.org/10.3847/2041-8213/abd401} {\bibfield  {journal} {\bibinfo
  {journal} {Astrophys. J. Lett.}\ }\textbf {\bibinfo {volume} {905}},\
  \bibinfo {pages} {L34} (\bibinfo {year} {2020})},\ \Eprint
  {https://arxiv.org/abs/2009.04496} {arXiv:2009.04496 [astro-ph.HE]}
  \BibitemShut {NoStop}%
\bibitem [{\citenamefont {Hobbs}\ \emph {et~al.}(2010)\citenamefont {Hobbs}
  \emph {et~al.}}]{Hobbs:2009yy}%
  \BibitemOpen
  \bibfield  {author} {\bibinfo {author} {\bibfnamefont {G.}~\bibnamefont
  {Hobbs}} \emph {et~al.},\ }\bibfield  {booktitle} {\emph {\bibinfo
  {booktitle} {{Gravitational waves. Proceedings, 8th Edoardo Amaldi
  Conference, Amaldi 8, New York, USA, June 22-26, 2009}}},\ }\href
  {https://doi.org/10.1088/0264-9381/27/8/084013} {\bibfield  {journal}
  {\bibinfo  {journal} {Class. Quant. Grav.}\ }\textbf {\bibinfo {volume}
  {27}},\ \bibinfo {pages} {084013} (\bibinfo {year} {2010})},\ \Eprint
  {https://arxiv.org/abs/0911.5206} {arXiv:0911.5206 [astro-ph.SR]}
  \BibitemShut {NoStop}%
\bibitem [{\citenamefont {Carilli}\ and\ \citenamefont
  {Rawlings}(2004)}]{Carilli:2004nx}%
  \BibitemOpen
  \bibfield  {author} {\bibinfo {author} {\bibfnamefont {C.~L.}\ \bibnamefont
  {Carilli}}\ and\ \bibinfo {author} {\bibfnamefont {S.}~\bibnamefont
  {Rawlings}},\ }\bibfield  {booktitle} {\emph {\bibinfo {booktitle}
  {{International SKA Conference 2003 Geraldton, Australia, July 27-August 2,
  2003}}},\ }\href {https://doi.org/10.1016/j.newar.2004.09.001} {\bibfield
  {journal} {\bibinfo  {journal} {New Astron. Rev.}\ }\textbf {\bibinfo
  {volume} {48}},\ \bibinfo {pages} {979} (\bibinfo {year} {2004})},\ \Eprint
  {https://arxiv.org/abs/astro-ph/0409274} {arXiv:astro-ph/0409274 [astro-ph]}
  \BibitemShut {NoStop}%
\bibitem [{\citenamefont {Amaro-Seoane}\ \emph {et~al.}(2017)\citenamefont
  {Amaro-Seoane} \emph {et~al.}}]{Audley:2017drz}%
  \BibitemOpen
  \bibfield  {author} {\bibinfo {author} {\bibfnamefont {P.}~\bibnamefont
  {Amaro-Seoane}} \emph {et~al.} (\bibinfo {collaboration} {LISA}),\
  }\href@noop {} {\  (\bibinfo {year} {2017})},\ \Eprint
  {https://arxiv.org/abs/1702.00786} {arXiv:1702.00786 [astro-ph.IM]}
  \BibitemShut {NoStop}%
\bibitem [{\citenamefont {Ruan}\ \emph {et~al.}(2018)\citenamefont {Ruan},
  \citenamefont {Guo}, \citenamefont {Cai},\ and\ \citenamefont
  {Zhang}}]{Guo:2018npi}%
  \BibitemOpen
  \bibfield  {author} {\bibinfo {author} {\bibfnamefont {W.-H.}\ \bibnamefont
  {Ruan}}, \bibinfo {author} {\bibfnamefont {Z.-K.}\ \bibnamefont {Guo}},
  \bibinfo {author} {\bibfnamefont {R.-G.}\ \bibnamefont {Cai}},\ and\ \bibinfo
  {author} {\bibfnamefont {Y.-Z.}\ \bibnamefont {Zhang}},\ }\href@noop {} {\
  (\bibinfo {year} {2018})},\ \Eprint {https://arxiv.org/abs/1807.09495}
  {arXiv:1807.09495 [gr-qc]} \BibitemShut {NoStop}%
\bibitem [{\citenamefont {Kawamura}\ \emph {et~al.}(2011)\citenamefont
  {Kawamura} \emph {et~al.}}]{Kawamura:2011zz}%
  \BibitemOpen
  \bibfield  {author} {\bibinfo {author} {\bibfnamefont {S.}~\bibnamefont
  {Kawamura}} \emph {et~al.},\ }\bibfield  {booktitle} {\emph {\bibinfo
  {booktitle} {{Laser interferometer space antenna. Proceedings, 8th
  International LISA Symposium, Stanford, USA, June 28-July 2, 2010}}},\ }\href
  {https://doi.org/10.1088/0264-9381/28/9/094011} {\bibfield  {journal}
  {\bibinfo  {journal} {Class. Quant. Grav.}\ }\textbf {\bibinfo {volume}
  {28}},\ \bibinfo {pages} {094011} (\bibinfo {year} {2011})}\BibitemShut
  {NoStop}%
\bibitem [{\citenamefont {Phinney}\ \emph {et~al.}(2004)\citenamefont
  {Phinney}, \citenamefont {Bender}, \citenamefont {Buchman}, \citenamefont
  {Byer}, \citenamefont {Cornish}, \citenamefont {Fritschel}, \citenamefont
  {Folkner}, \citenamefont {Merkowitz}, \citenamefont {Danzmann}, \citenamefont
  {DiFiore} \emph {et~al.}}]{phinney2004big}%
  \BibitemOpen
  \bibfield  {author} {\bibinfo {author} {\bibfnamefont {S.}~\bibnamefont
  {Phinney}}, \bibinfo {author} {\bibfnamefont {P.}~\bibnamefont {Bender}},
  \bibinfo {author} {\bibfnamefont {R.}~\bibnamefont {Buchman}}, \bibinfo
  {author} {\bibfnamefont {R.}~\bibnamefont {Byer}}, \bibinfo {author}
  {\bibfnamefont {N.}~\bibnamefont {Cornish}}, \bibinfo {author} {\bibfnamefont
  {P.}~\bibnamefont {Fritschel}}, \bibinfo {author} {\bibfnamefont
  {W.}~\bibnamefont {Folkner}}, \bibinfo {author} {\bibfnamefont
  {S.}~\bibnamefont {Merkowitz}}, \bibinfo {author} {\bibfnamefont
  {K.}~\bibnamefont {Danzmann}}, \bibinfo {author} {\bibfnamefont
  {L.}~\bibnamefont {DiFiore}}, \emph {et~al.},\ }\href@noop {} {\bibfield
  {journal} {\bibinfo  {journal} {NASA Mission Concept Study}\ } (\bibinfo
  {year} {2004})}\BibitemShut {NoStop}%
\bibitem [{\citenamefont {Aasi}\ \emph {et~al.}(2015)\citenamefont {Aasi} \emph
  {et~al.}}]{TheLIGOScientific:2014jea}%
  \BibitemOpen
  \bibfield  {author} {\bibinfo {author} {\bibfnamefont {J.}~\bibnamefont
  {Aasi}} \emph {et~al.} (\bibinfo {collaboration} {LIGO Scientific}),\ }\href
  {https://doi.org/10.1088/0264-9381/32/7/074001} {\bibfield  {journal}
  {\bibinfo  {journal} {Class. Quant. Grav.}\ }\textbf {\bibinfo {volume}
  {32}},\ \bibinfo {pages} {074001} (\bibinfo {year} {2015})},\ \Eprint
  {https://arxiv.org/abs/1411.4547} {arXiv:1411.4547 [gr-qc]} \BibitemShut
  {NoStop}%
\bibitem [{\citenamefont {Somiya}(2012)}]{Somiya:2011np}%
  \BibitemOpen
  \bibfield  {author} {\bibinfo {author} {\bibfnamefont {K.}~\bibnamefont
  {Somiya}} (\bibinfo {collaboration} {KAGRA}),\ }\bibfield  {booktitle} {\emph
  {\bibinfo {booktitle} {{Gravitational waves. Numerical relativity - data
  analysis. Proceedings, 9th Edoardo Amaldi Conference, Amaldi 9, and meeting,
  NRDA 2011, Cardiff, UK, July 10-15, 2011}}},\ }\href
  {https://doi.org/10.1088/0264-9381/29/12/124007} {\bibfield  {journal}
  {\bibinfo  {journal} {Class. Quant. Grav.}\ }\textbf {\bibinfo {volume}
  {29}},\ \bibinfo {pages} {124007} (\bibinfo {year} {2012})},\ \Eprint
  {https://arxiv.org/abs/1111.7185} {arXiv:1111.7185 [gr-qc]} \BibitemShut
  {NoStop}%
\bibitem [{\citenamefont {Reitze}\ \emph {et~al.}(2019)\citenamefont {Reitze}
  \emph {et~al.}}]{Reitze:2019iox}%
  \BibitemOpen
  \bibfield  {author} {\bibinfo {author} {\bibfnamefont {D.}~\bibnamefont
  {Reitze}} \emph {et~al.},\ }\href@noop {} {\bibfield  {journal} {\bibinfo
  {journal} {Bull. Am. Astron. Soc.}\ }\textbf {\bibinfo {volume} {51}},\
  \bibinfo {pages} {035} (\bibinfo {year} {2019})},\ \Eprint
  {https://arxiv.org/abs/1907.04833} {arXiv:1907.04833 [astro-ph.IM]}
  \BibitemShut {NoStop}%
\bibitem [{\citenamefont {Punturo}\ \emph {et~al.}(2010)\citenamefont {Punturo}
  \emph {et~al.}}]{Punturo:2010zz}%
  \BibitemOpen
  \bibfield  {author} {\bibinfo {author} {\bibfnamefont {M.}~\bibnamefont
  {Punturo}} \emph {et~al.},\ }\bibfield  {booktitle} {\emph {\bibinfo
  {booktitle} {{Proceedings, 14th Workshop on Gravitational wave data analysis
  (GWDAW-14): Rome, Italy, January 26-29, 2010}}},\ }\href
  {https://doi.org/10.1088/0264-9381/27/19/194002} {\bibfield  {journal}
  {\bibinfo  {journal} {Class. Quant. Grav.}\ }\textbf {\bibinfo {volume}
  {27}},\ \bibinfo {pages} {194002} (\bibinfo {year} {2010})}\BibitemShut
  {NoStop}%
\bibitem [{\citenamefont {Schmitz}(2021)}]{Schmitz:2020syl}%
  \BibitemOpen
  \bibfield  {author} {\bibinfo {author} {\bibfnamefont {K.}~\bibnamefont
  {Schmitz}},\ }\href {https://doi.org/10.1007/JHEP01(2021)097} {\bibfield
  {journal} {\bibinfo  {journal} {JHEP}\ }\textbf {\bibinfo {volume} {01}},\
  \bibinfo {pages} {097}},\ \Eprint {https://arxiv.org/abs/2002.04615}
  {arXiv:2002.04615 [hep-ph]} \BibitemShut {NoStop}%
\bibitem [{\citenamefont {Aghanim}\ \emph {et~al.}(2020)\citenamefont {Aghanim}
  \emph {et~al.}}]{Planck:2018vyg}%
  \BibitemOpen
  \bibfield  {author} {\bibinfo {author} {\bibfnamefont {N.}~\bibnamefont
  {Aghanim}} \emph {et~al.} (\bibinfo {collaboration} {Planck}),\ }\href
  {https://doi.org/10.1051/0004-6361/201833910} {\bibfield  {journal} {\bibinfo
   {journal} {Astron. Astrophys.}\ }\textbf {\bibinfo {volume} {641}},\
  \bibinfo {pages} {A6} (\bibinfo {year} {2020})},\ \bibinfo {note} {[Erratum:
  Astron.Astrophys. 652, C4 (2021)]},\ \Eprint
  {https://arxiv.org/abs/1807.06209} {arXiv:1807.06209 [astro-ph.CO]}
  \BibitemShut {NoStop}%
\bibitem [{\citenamefont {Ashoorioon}(2015)}]{Ashoorioon:2015hya}%
  \BibitemOpen
  \bibfield  {author} {\bibinfo {author} {\bibfnamefont {A.}~\bibnamefont
  {Ashoorioon}},\ }\href {https://doi.org/10.1016/j.physletb.2015.06.022}
  {\bibfield  {journal} {\bibinfo  {journal} {Phys. Lett. B}\ }\textbf
  {\bibinfo {volume} {747}},\ \bibinfo {pages} {446} (\bibinfo {year}
  {2015})},\ \Eprint {https://arxiv.org/abs/1502.00556} {arXiv:1502.00556
  [astro-ph.CO]} \BibitemShut {NoStop}%
\bibitem [{\citenamefont {Ashoorioon}\ \emph {et~al.}(2021)\citenamefont
  {Ashoorioon}, \citenamefont {Rostami},\ and\ \citenamefont
  {Firouzjaee}}]{Ashoorioon:2020hln}%
  \BibitemOpen
  \bibfield  {author} {\bibinfo {author} {\bibfnamefont {A.}~\bibnamefont
  {Ashoorioon}}, \bibinfo {author} {\bibfnamefont {A.}~\bibnamefont
  {Rostami}},\ and\ \bibinfo {author} {\bibfnamefont {J.~T.}\ \bibnamefont
  {Firouzjaee}},\ }\href {https://doi.org/10.1103/PhysRevD.103.123512}
  {\bibfield  {journal} {\bibinfo  {journal} {Phys. Rev. D}\ }\textbf {\bibinfo
  {volume} {103}},\ \bibinfo {pages} {123512} (\bibinfo {year} {2021})},\
  \Eprint {https://arxiv.org/abs/2012.02817} {arXiv:2012.02817 [astro-ph.CO]}
  \BibitemShut {NoStop}%
\bibitem [{\citenamefont {Deng}\ and\ \citenamefont
  {Vilenkin}(2017)}]{Deng:2017uwc}%
  \BibitemOpen
  \bibfield  {author} {\bibinfo {author} {\bibfnamefont {H.}~\bibnamefont
  {Deng}}\ and\ \bibinfo {author} {\bibfnamefont {A.}~\bibnamefont
  {Vilenkin}},\ }\href {https://doi.org/10.1088/1475-7516/2017/12/044}
  {\bibfield  {journal} {\bibinfo  {journal} {JCAP}\ }\textbf {\bibinfo
  {volume} {12}},\ \bibinfo {pages} {044}},\ \Eprint
  {https://arxiv.org/abs/1710.02865} {arXiv:1710.02865 [gr-qc]} \BibitemShut
  {NoStop}%
\bibitem [{\citenamefont {Deng}(2020)}]{Deng:2020mds}%
  \BibitemOpen
  \bibfield  {author} {\bibinfo {author} {\bibfnamefont {H.}~\bibnamefont
  {Deng}},\ }\href {https://doi.org/10.1088/1475-7516/2020/09/023} {\bibfield
  {journal} {\bibinfo  {journal} {JCAP}\ }\textbf {\bibinfo {volume} {09}},\
  \bibinfo {pages} {023}},\ \Eprint {https://arxiv.org/abs/2006.11907}
  {arXiv:2006.11907 [astro-ph.CO]} \BibitemShut {NoStop}%
\bibitem [{\citenamefont {Papanikolaou}\ \emph {et~al.}(2021)\citenamefont
  {Papanikolaou}, \citenamefont {Vennin},\ and\ \citenamefont
  {Langlois}}]{Papanikolaou:2020qtd}%
  \BibitemOpen
  \bibfield  {author} {\bibinfo {author} {\bibfnamefont {T.}~\bibnamefont
  {Papanikolaou}}, \bibinfo {author} {\bibfnamefont {V.}~\bibnamefont
  {Vennin}},\ and\ \bibinfo {author} {\bibfnamefont {D.}~\bibnamefont
  {Langlois}},\ }\href {https://doi.org/10.1088/1475-7516/2021/03/053}
  {\bibfield  {journal} {\bibinfo  {journal} {JCAP}\ }\textbf {\bibinfo
  {volume} {03}},\ \bibinfo {pages} {053}},\ \Eprint
  {https://arxiv.org/abs/2010.11573} {arXiv:2010.11573 [astro-ph.CO]}
  \BibitemShut {NoStop}%
\bibitem [{\citenamefont {Papanikolaou}\ \emph {et~al.}(2022)\citenamefont
  {Papanikolaou}, \citenamefont {Tzerefos}, \citenamefont {Basilakos},\ and\
  \citenamefont {Saridakis}}]{Papanikolaou:2021uhe}%
  \BibitemOpen
  \bibfield  {author} {\bibinfo {author} {\bibfnamefont {T.}~\bibnamefont
  {Papanikolaou}}, \bibinfo {author} {\bibfnamefont {C.}~\bibnamefont
  {Tzerefos}}, \bibinfo {author} {\bibfnamefont {S.}~\bibnamefont
  {Basilakos}},\ and\ \bibinfo {author} {\bibfnamefont {E.~N.}\ \bibnamefont
  {Saridakis}},\ }\href {https://doi.org/10.1088/1475-7516/2022/10/013}
  {\bibfield  {journal} {\bibinfo  {journal} {JCAP}\ }\textbf {\bibinfo
  {volume} {10}},\ \bibinfo {pages} {013}},\ \Eprint
  {https://arxiv.org/abs/2112.15059} {arXiv:2112.15059 [astro-ph.CO]}
  \BibitemShut {NoStop}%
\bibitem [{\citenamefont {Bhaumik}\ \emph
  {et~al.}(2022{\natexlab{a}})\citenamefont {Bhaumik}, \citenamefont
  {Ghoshal},\ and\ \citenamefont {Lewicki}}]{Bhaumik:2022pil}%
  \BibitemOpen
  \bibfield  {author} {\bibinfo {author} {\bibfnamefont {N.}~\bibnamefont
  {Bhaumik}}, \bibinfo {author} {\bibfnamefont {A.}~\bibnamefont {Ghoshal}},\
  and\ \bibinfo {author} {\bibfnamefont {M.}~\bibnamefont {Lewicki}},\ }\href
  {https://doi.org/10.1007/JHEP07(2022)130} {\bibfield  {journal} {\bibinfo
  {journal} {JHEP}\ }\textbf {\bibinfo {volume} {07}},\ \bibinfo {pages}
  {130}},\ \Eprint {https://arxiv.org/abs/2205.06260} {arXiv:2205.06260
  [astro-ph.CO]} \BibitemShut {NoStop}%
\bibitem [{\citenamefont {Bhaumik}\ \emph
  {et~al.}(2022{\natexlab{b}})\citenamefont {Bhaumik}, \citenamefont {Ghoshal},
  \citenamefont {Jain},\ and\ \citenamefont {Lewicki}}]{Bhaumik:2022zdd}%
  \BibitemOpen
  \bibfield  {author} {\bibinfo {author} {\bibfnamefont {N.}~\bibnamefont
  {Bhaumik}}, \bibinfo {author} {\bibfnamefont {A.}~\bibnamefont {Ghoshal}},
  \bibinfo {author} {\bibfnamefont {R.~K.}\ \bibnamefont {Jain}},\ and\
  \bibinfo {author} {\bibfnamefont {M.}~\bibnamefont {Lewicki}},\ }\href@noop
  {} {\  (\bibinfo {year} {2022}{\natexlab{b}})},\ \Eprint
  {https://arxiv.org/abs/2212.00775} {arXiv:2212.00775 [astro-ph.CO]}
  \BibitemShut {NoStop}%
\bibitem [{\citenamefont {Dom\`enech}\ \emph {et~al.}(2021)\citenamefont
  {Dom\`enech}, \citenamefont {Lin},\ and\ \citenamefont
  {Sasaki}}]{Domenech:2020ssp}%
  \BibitemOpen
  \bibfield  {author} {\bibinfo {author} {\bibfnamefont {G.}~\bibnamefont
  {Dom\`enech}}, \bibinfo {author} {\bibfnamefont {C.}~\bibnamefont {Lin}},\
  and\ \bibinfo {author} {\bibfnamefont {M.}~\bibnamefont {Sasaki}},\ }\href
  {https://doi.org/10.1088/1475-7516/2021/11/E01} {\bibfield  {journal}
  {\bibinfo  {journal} {JCAP}\ }\textbf {\bibinfo {volume} {04}},\ \bibinfo
  {pages} {062}},\ \bibinfo {note} {[Erratum: JCAP 11, E01 (2021)]},\ \Eprint
  {https://arxiv.org/abs/2012.08151} {arXiv:2012.08151 [gr-qc]} \BibitemShut
  {NoStop}%
\bibitem [{\citenamefont {Liu}\ \emph {et~al.}(2022)\citenamefont {Liu},
  \citenamefont {Bian}, \citenamefont {Cai}, \citenamefont {Guo},\ and\
  \citenamefont {Wang}}]{Liu:2022lvz}%
  \BibitemOpen
  \bibfield  {author} {\bibinfo {author} {\bibfnamefont {J.}~\bibnamefont
  {Liu}}, \bibinfo {author} {\bibfnamefont {L.}~\bibnamefont {Bian}}, \bibinfo
  {author} {\bibfnamefont {R.-G.}\ \bibnamefont {Cai}}, \bibinfo {author}
  {\bibfnamefont {Z.-K.}\ \bibnamefont {Guo}},\ and\ \bibinfo {author}
  {\bibfnamefont {S.-J.}\ \bibnamefont {Wang}},\ }\href@noop {} {\  (\bibinfo
  {year} {2022})},\ \Eprint {https://arxiv.org/abs/2208.14086}
  {arXiv:2208.14086 [astro-ph.CO]} \BibitemShut {NoStop}%
\end{thebibliography}%
	
\end{document}